\documentclass{JHEP3}
\usepackage{latexsym}
\usepackage{amsmath}
\usepackage{graphicx}

\title{Glueball Scattering Amplitudes from Holography}

\author{Wolfgang M\"uck \\
        Dipartimento di Scienze Fisiche,
	Universit\`a di Napoli ``Federico II'' and \\
	I.N.F.N.\ -- Sezione di Napoli, \\
	Via Cintia, 80126 Napoli, Italy \\ 
	E-mail: \email{mueck@na.infn.it}} 

\author{Maurizio Prisco\\ 
        Dipartimento di Fisica, Universit\`a di Roma ``Tor
        Vergata'' and \\
        I.N.F.N.\ -- Sezione di Roma ``Tor Vergata'', \\
        Via della Ricerca Scientifica 1, 00133  Roma,  Italy\\
        E-mail: \email{maurizio.prisco@roma2.infn.it}} 

\abstract{Using techniques developed in a previous paper
  three-point functions in field theories described by holographic
  renormalization group flows are computed. We consider a system of
  one active scalar and one inert scalar coupled to gravity. For the
  GPPZ flow, their dual operators create states that are 
  interpreted as glueballs of the $\mathcal{N}=1$ SYM theory, which
  lies at the infrared end of the renormalization group flow. The
  scattering amplitudes for three-glueball processes are
  calculated providing precise predictions for glueball decays in 
  $\mathcal{N}=1$ SYM theory. Numerical results for low-lying
  glueballs are included.} 

\preprint{ROM2F/2004/03 \\NA-DSF-2-2004 \\hep-th/0402068}


\providecommand{\ie}{\emph{i.e.,\ }}
\providecommand{\eg}{\emph{e.g.,\ }}

\providecommand{\rmd}{\mathrm{d}} 

\providecommand{\dr}{\rmd r}
\providecommand{\e}[1]{\mathrm{e}^{#1}}

\providecommand{\F}{\mathrm{F}}

\providecommand{\LeP}{\mathrm{P}}

\providecommand{\K}{\mathcal{K}}

\providecommand{\tg}{\tilde{g}}

\providecommand{\bp}{\bar{\phi}}
\providecommand{\bs}{\bar{\sigma}}

\providecommand{\hp}{\hat{\varphi}}

\providecommand{\ha}{\hat{a}}

\providecommand{\hpsi}{\hat{\psi}}
\providecommand{\rpsi}{\check{\psi}}

\providecommand{\htt}{{h^{TT}}}

\providecommand{\ta}{\tilde{a}}

\providecommand{\tK}{\tilde{K}}

\providecommand{\vev}[1]{\left\langle #1 \right\rangle}

\providecommand{\X}{\mathcal{X}}
\providecommand{\Y}{\mathcal{Y}}

\providecommand{\Op}{\mathcal{O}}

\providecommand{\tr}{\mathrm{tr}\,}   

\begin{document}


\section{Introduction}
It is a paradigm of the AdS/CFT correspondence that the dynamics of
a field theory living in an (asymptotically) anti-de Sitter bulk
space-time encodes the correlation functions of its dual (deformed)
conformal field theory \cite{Gubser:1998bc,Witten:1998qj}. 
The correspondence holds \emph{par excellence} for $\mathcal{N}=8$
gauged supergravity in $(d+1)=5$ dimensions on the bulk side, which is
obtained from $D=10$ type IIB supergravity by compactification on a
five-sphere. Its $AdS_5$ solution is dual, in the planar limit and at
strong 't~Hooft coupling, to $d=4$, $\mathcal{N}=4$ super Yang-Mills
(SYM) theory, which is a conformal quantum field theory
\cite{Maldacena:1998re}.

The study of fluctuations around other bulk configurations, which are
interpreted as the duals of Renormalization Group (RG) flows of
$\mathcal{N}=4$ SYM theory driven by relevant operators, is
interesting, because it yields the correlation functions of the
respective dual quantum field theories in a regime that is not
accessible by ordinary perturbation theory. For example, the GPPZ flow
\cite{Girardello:1999bd}, which is a supersymmetric mass deformation
of $\mathcal{N}=4$ SYM theory, possesses many qualitative features of
pure $\mathcal{N}=1$ SYM theory, in particular quark
confinement.\footnote{An essential feature of pure $\mathcal{N}=1$ SYM
  theory is missing in the GPPZ flow, namely the gaugino
  condensate. There exists a family of analytic solutions of bulk
  backgrounds, which include a non-zero gaugino condensate
  \cite{Girardello:1999bd}, but the linearized fluctuation equations
  around these backgrounds are not analytically solvable, so that one
  must resort to numerical methods to obtain, \eg the mass spectrum of
  states \cite{Apreda:2003sy}.}    
The two-point functions in the GPPZ flow, which have been calculated
using holography, exhibit a spectrum of particles, which must arise
from as yet unknown non-perturbative effects in the field
theory. These particles are interpreted as glueballs of the 
$\mathcal{N}=1$ SYM theory, which lies at the infra-red end of 
the RG flow \cite{Bianchi:2000sm}. Naturally, a calculation of the
glueball scattering amplitudes would be very desirable.

Until recently, the holographic calculation of $n$-point functions with
$n>2$ involving the operators that are dual to the active
scalars\footnote{In the  
  common nomenclature, an \emph{active} scalar is dual to the 
  operator driving the RG flow and has a non-zero background value,
  whereas the other scalars are called \emph{inert}.} 
of holographic RG flow backgrounds was considered unfeasible, because
their fluctuations couple to the fluctuations of the bulk metric even
at the linearized level. In contrast, inert scalars do not couple to
the bulk metric at the linearized level, and the calculation of the
corresponding three-point functions is rather straightforward
\cite{Bianchi:2003bd}. In a recent paper
\cite{Bianchi:2003ug}, however, enormous progress was made exploiting
a gauge invariant formalism, in which the true degrees of freedom
of the bulk metric decouple from the active scalar fluctuations at the
linearized level. Thus, this formalism simplifies the calculations of
two-point functions for the active scalars and makes higher $n$-point
functions accessible by the use of the Green's function method. 
This was demonstrated for the case of a single active scalar, which is
dual to an operator $\Op$, by calculating the three-point function
$\vev{\Op \Op \Op}$. An important result of
\cite{Bianchi:2003ug} was the proof of Bose symmetry, which was not
obvious from the formal expression resulting from the Green's function
method. This proof involved two steps, which we shall summarize
here. In the first step a field redefinition was used in order to
eliminate those terms from the quadratic source of the field equations
that contain two radial derivatives. Then, after using momentum
conservation, some terms in the radial integral were integrated by
parts, and the final, Bose symmetric result was obtained.
Moreover, it was shown (for the GPPZ flow) that the boundary terms
from the integration by parts cancel the contribution from
the field redefinition. 

In this paper, we continue the programme of \cite{Bianchi:2003ug}
considering a system of one active and one inert scalar coupled to
bulk gravity in a generic holographic RG flow background and finding
the expressions for all non-local three-point functions of the
respective dual operators. Applied to the GPPZ flow, the results are then
used to calculate the scattering amplitudes for the glueballs
generated by these operators. Thus, for the first time, we are able to
predict glueball scattering amplitudes for an $\mathcal{N}=1$ SYM
theory using holography.\footnote{The qualitative behaviour of
  glueball scattering in confining gauge theories with string duals
  has been discussed before, see the recent talk 
  \cite{Boschi-Filho:2003ej} and references therein.}

In order to avoid being repetitive we chose not to draft this paper in
a self-contained fashion, but to make essential use of the material
presented in \cite{Bianchi:2003ug}. Hence, we urge the reader to
consult that paper first in order to become familiar with the gauge
invariant method and our notation. We also refer to
\cite{Bianchi:2003ug} for more references to the relevant literature.

The rest of this paper is organized as follows. In Sec.~\ref{field_eq}
we review the equations of motion using the gauge invariant formalism
developed in \cite{Bianchi:2003ug}. Sec.~\ref{Sec_threepoint}, which
contains the main achievements of our work, is
devoted to the study of three-point functions in a generic holographic
RG flow background containing one active and one inert scalar
field. Together with the bulk graviton, the three respective dual
boundary operators give rise to a total of ten independent three-point
functions, all of which will be calculated. We proceed as in
\cite{Bianchi:2003ug} performing first a field redefinition and then a
suitable integration by parts in order to render the final results
Bose symmetric. However, we shall not be concerned about whether the
boundary terms resulting from the two steps cancel each other,
as the net boundary terms would contribute only contact terms to the
three-point functions. Our final expressions for the
three-point functions have the form of generically divergent
integrals. The divergences, which have the form of contact terms, 
can be easily understood and removed on a case-by-case basis by
comparison with the results of holographic renormalization, and, thus,
we shall not do it explicitly for the general setup. Such an attitude
is justified, because contact terms have no effect on the scattering
amplitudes.

As an application of our results of Sec.~\ref{Sec_threepoint} we shall
analyze the GPPZ flow in Secs.~\ref{GPPZ_corrs} and \ref{scatt}. In
Sec.~\ref{GPPZ_corrs}, we start by repeating the 
analysis of the bulk-to-boundary propagators, which give rise to the
two-point functions. Then, we shall present the expressions that
encode the non-local three-point functions and scattering amplitudes. 
Moreover, we shall illustrate how the divergences of the three-point
function integrals are understood and removed by using holographic
renormalization. Finally, in Sec.~\ref{scatt}, we discuss some physical
interpretations of the calculated scattering amplitudes, in particular
the possible glueball decay channels. Some useful
relations for the GPPZ flow can be found in appendix~\ref{GPPZ_rels},
and numerical results for the decay amplitudes are listed in
appendix~\ref{decay_amps}. 

Let us stress that we are interested only in the non-local three-point
functions. A detailed analysis of the contact terms would be a
worthwhile exercise, because they play a crucial role in the
consistency of the subtraction procedure \cite{Bianchi:2001kw} and
contribute to certain sum rules \cite{Anselmi:2002fk}. Such an
analysis would have to take into account not only the contact terms
that result from holographic renormalization
\cite{deHaro:2000xn,Bianchi:2001kw,Martelli:2002sp}, but also the use
of the gauge invariant fields, the field redefinition and the
integration by parts, which occur in our calculation. In particular,
analyzing the contributions from the use of the gauge invariant fields
might be rather challenging, because one would need to translate the
fields into the axial gauge, in which holographic renormalization has been
carried out. Moreover, this translation must be done to third order in the
fluctuations (to obtain all local terms in the three-point functions),
but we have only derived the linear relations, which are sufficient
for the analysis of the bulk equations of motion. Alternatively, one
might try to perform holographic renormalization in a gauge invariant
fashion. 
However, contact terms can be calculated only on a case-by-case
basis, \eg for the GPPZ flow. Thus, it would be impossible to include
them into our general results of Sec.~\ref{Sec_threepoint}. Having in
mind to calculate the scattering amplitudes for on-shell glueballs in
the GPPZ flow, we feel justified in omitting them completely.


\section{Field Equations}
\label{field_eq}
In this section, we shall present the equations of motion to quadratic
order in the fluctuations for a bulk system 
containing one active scalar, $\phi$, and one inert scalar, $\sigma$,
coupled to gravity. Our main reason for this restriction of the number
of scalars is that, generically, it is impossible to diagonalize the
effective mass terms of scalars of the same kind (active or inert) in
holographic RG flow backgrounds. Thus, we ensure that the
\emph{linear} second order ODEs for the true degrees of 
freedom are not coupled, which makes the use of the Green's function method
straightforward. Our results are easily generalized to the
case of several inert scalars with diagonal effective mass terms. 

The equations of motion shall be presented in the gauge invariant
approach of \cite{Bianchi:2003ug}. In this approach, the field
fluctuations are combined into a set of gauge invariant variables, in
terms of which the equations of motion are expressed. The remaining
fluctuations describe gauge artifacts and are explicitly dropped. 

For completeness, we shall review in subsection \ref{background} the
general relations defining a holographic RG flow background and the
definitions of the gauge invariant variables. 
Then, in subsections \ref{active_eq}, \ref{inert_eq} and
\ref{grav_eq}, respectively, the equations of motion for the active
scalar, the inert scalar and the traceless transversal parts of the
bulk metric fluctuations are presented.

\subsection{Preliminaries}
\label{background}
A generic holographic RG flow background with one active scalar
satisfies \cite{Freedman:1999gp,Skenderis:1999mm,DeWolfe:1999cp} 
\begin{equation}
\label{background_eq}
\begin{split}
  \rmd s^2 &= \rmd r^2 + \e{2A(r)} \eta_{ij} \rmd x^i \rmd x^j~,\\
  \partial_r A(r) &= -\frac{2}{d-1} W(\bp)~, \\
  \partial_r \bp &= W_\phi(\bp)~, \quad
  \bs =0~,
\end{split}
\end{equation}
where the superpotential $W$ is a function of the active scalar and
satisfies the following functional equation, which is imposed by the
supersymmetry of the background,
\begin{equation}
\label{Wdef}
  \frac12 W_\phi^2 -\frac{d}{d-1} W^2 = V|_{\sigma=0}~.
\end{equation}
Here and henceforth, derivatives with respect to the fields are
denoted by subscripts, as in $W_\phi= \rmd W/\rmd \phi$.
Notice that, in general, \eqref{Wdef} does not include the inert
scalar, although in some cases (\eg for the GPPZ flow) 
it might be possible that the full potential $V$ is expressible in
terms of a suitable superpotential containing also the inert scalar. 

The couplings between the inert scalar and the active scalar are
strongly restricted by the fact that the relation $V_\sigma=
\partial V/\partial \sigma = 0$ holds in the background 
due to the equation of motion for $\sigma$. Taking a derivative of
this relation with respect to $r$ we find 
\begin{equation}
\label{drVI}
  0= \partial_r V_\sigma = V_{\sigma\phi} W_\phi~,
\end{equation}
so that we find also $V_{\sigma\phi}=0$. Applying this argument recursively
one arrives at the conclusion that 
\begin{equation}
\label{VIphi}
   V_{\sigma\phi\cdots\phi} =0
\end{equation}
for any number of $\phi$-derivatives.
Notice that this argument does not hold in the more general case with
several active scalars. In that case the right hand side of
\eqref{drVI} contains a sum over all active scalars, and the
members of the sum need not be zero individually. 

We shall now review the definition of the gauge invariant variables. 
The bulk system is treated in the time slicing formalism, where the bulk
metric is written in the form
\begin{equation} 
\label{tslice_metric} 
  \rmd s^2 = (n^2+n_in^i) \dr^2 + 2 n_i \rmd r \rmd x^i  
             + g_{ij} \rmd x^i \rmd x^j~. 
\end{equation} 
The fluctuations of the various fields around the background
\eqref{background_eq} are introduced by 
\begin{equation} 
\label{expand} 
\begin{aligned} 
  \phi   &= \bp(r) +\varphi~, \qquad &\sigma &= \sigma,\\ 
  n_i    &= \nu_i~, &                    n   &= 1+ \nu~,\\ 
  g_{ij} &= \e{2A(r)} \left( \eta_{ij} + h_{ij} \right)~. 
\end{aligned} 
\end{equation} 
Furthermore, the metric fluctuations $h^i_j$ are split as
follows,
\begin{equation} 
\label{hsplit} 
 h^i_j = \htt^i_j  
  + \partial^i \epsilon_j +\partial_j \epsilon^i 
  + \frac{\partial^i \partial_j}{\Box} H + \frac1{d-1} \delta^i_j h~, 
\end{equation} 
where $\htt^i_j$ denotes the traceless transversal part, and 
$\epsilon^i$ is a transversal vector ($\partial_i \epsilon^i=0$).

The inert scalar, $\sigma$, is gauge invariant to lowest
order. Therefore, we shall use the same symbol to denote the
corresponding (all order) gauge invariant variable. 
The remaining first-order gauge invariant combinations of the
fluctuations are 
\begin{align} 
\label{A} 
 a &= \varphi +W_\phi \frac{h}{4W}~,\\ 
\label{B} 
 b &= \nu + \partial_r \left( \frac{h}{4W} \right)~,\\ 
\label{C} 
 c &= \partial_i \nu^i + \Box \frac{h}{4W}  
 -\frac12 \e{2A} \partial_r H~,\\  
\label{D} 
 d^i &= \Pi^i_j \nu^j - \e{2A} \partial_r \epsilon^i~, \phantom{\frac12}\\ 
\label{E}  
 e^i_j &= h^{TT}{}^i_j = \Pi^{ik}_{jl} h^l_k~. 
\end{align}
Here and henceforth, $\Pi^i_j$ and $\Pi^{ik}_{jl}$ denote the
transversal and the traceless transversal projectors, respectively,
\begin{align}
\label{Pi_ij_def}
  \Pi^i_j &= \delta^i_j -\frac{\partial^i \partial_j}{\Box}~,\\
\label{TTproj_def}
  \Pi^{ik}_{jl} &= \frac12 \left(\Pi^{ik} \Pi_{jl}  + \Pi^i_l \Pi^k_j
  \right) - \frac1{d-1} \Pi^i_j \Pi^k_l~.
\end{align}
The gauge invariant approach is embodied in a simple recipe, according
to which one expands the field equations to the desired order and 
replaces the fluctuations as follows, 
\begin{equation} 
\label{field_subs} 
  \varphi  \to a~, \quad 
  \nu      \to b~, \quad
  \nu^i    \to d^i + \frac{\partial^i c}{\Box}~, \quad
  h^i_j \to e^i_j~. 
\end{equation} 
The fluctuations $\epsilon^i$, $h$ and $H$ represent gauge artifacts
and are explicitly dropped, while the inert scalar $\sigma$ remains
unchanged. 

Finally we mention that, as in \cite{Bianchi:2003ug}, we raise and
lower the indices of 
fluctuations and of partial derivatives using the flat metric.

\subsection{The Equation for the Active Scalar}
\label{active_eq}
The equations of motion for the active scalar can be taken over from
Sec.~4 of \cite{Bianchi:2003ug} with only minor modifications, which
come in the form of additional terms that are quadratic in the inert
scalar. Terms with only one $\sigma$ cannot
occur because of \eqref{VIphi}.
For completeness, we repeat the full expressions here.  
First, the scalar equation remains
\begin{equation} 
\label{eq_a} 
  \left( \partial_r^2 -\frac{2d}{d-1} W \partial_r +\e{-2A} \Box 
  -V_{\phi\phi} \right) a -W_\phi \e{-2A} c -W_\phi \partial_r b 
  -2 V_\phi b = J_a~, 
\end{equation} 
where the source $J_a$ is now given by
\begin{equation} 
\label{J_a} 
\begin{split} 
  J_a &=  
  \frac12 V_{\phi\phi\phi} a^2 + V_\phi b^2 + 2 V_{\phi\phi} ab 
  - W_\phi b \partial_r b + (\partial_r a)(\partial_r b)  
  + \frac12 W_\phi e^i_j \partial_r e^j_i \\ 
  &\quad +\e{-2A} \left[ -2 b\Box a -(\partial^i b)(\partial_i a)  
  + c \partial_r a + 2\left( d^i +\frac{\partial^i c}{\Box} \right)  
  \partial_i \partial_r a  \right.\\ 
  &\quad - W_\phi \left(d^i +\frac{\partial^i c}{\Box} \right) 
\partial_i b  
  + \left( \partial_r  d^i + \partial_r \frac{\partial^i c}{\Box} 
\right) \partial_i a  
  - 2\frac{d-2}{d-1}W \left(d^i +\frac{\partial^i c}{\Box} \right) 
  \partial_i a \\ 
  &\left. \phantom{\frac12}  
  + e^i_j \left(\partial_i \partial^j a -W_\phi \partial_i d^j 
  -W_\phi \frac{\partial_i  \partial^j}{\Box} c\right) \right] 
  +\frac12 V_{\phi \sigma \sigma} \sigma^2~. 
\end{split}  
\end{equation} 
Second, the normal component of Einstein's equation is
\begin{equation} 
\label{eq_c} 
  -4W \e{-2A} c + 4 W_\phi \partial_r a -4 V_\phi a -8 V b =J_c~, 
\end{equation} 
where the source $J_c$ is easily generalized from (4.6) of
\cite{Bianchi:2003ug},
\begin{equation} 
\label{J_c} 
\begin{split} 
  J_c &= 4V b^2 +8 V_\phi ab +2 V_{\phi\phi} a^2 - 2 (\partial_r a)^2   
  + (\e{-2A} c)^2 + 2 \e{-2A} (\partial^i a)(\partial_i a) \\ 
  &\quad 
  + 4 W_\phi \e{-2A} \left(d^i +\frac{\partial^i c}{\Box} \right) 
  \partial_i a 
  +2 W e^i_j \partial_r e^j_i  
  -4 W \e{-2A} e^i_j \partial_i  
  \left(d^j +\frac{\partial^j c}{\Box} \right) \\ 
  &\quad 
  -\frac14 (\partial_r e^i_j)(\partial_r e^j_i)  
  +\e{-2A} \left( \partial_i d^j +\frac{\partial_i\partial^j}{\Box}
  c\right) \partial_r e^i_j \\ 
  &\quad -\e{-4A} \left[ \frac12 (\partial_i d^j)(\partial^i d_j)
  +\frac12 (\partial_i d^j)(\partial_j d^i) +2 (\partial_i d^j)  
  \frac{\partial^i\partial_j}{\Box} c +
  \left(\frac{\partial_i \partial^j}{\Box} c \right) 
  \left(\frac{\partial^i \partial_j}{\Box} c \right)  \right] \\ 
  &\quad -\e{-2A} \left[e^i_j  \Box e^j_i 
  +\frac34 (\partial_i e^j_k)(\partial^i e_j^k)  
  -\frac12 (\partial_i e^j_k)(\partial^k e^i_j) 
  \right]\\
  &\quad +2V_{\sigma\sigma} \sigma^2 
  -2 (\partial_r \sigma)(\partial_r \sigma) 
  +2\e{-2A} (\partial^i \sigma)(\partial_i \sigma)~. 
\end{split} 
\end{equation} 
Third, the mixed components of Einstein's equation yield
\begin{equation} 
\label{eq_bd} 
  -\frac12 \e{-2A} \Box d_i -2 W \partial_i b -2W_\phi \partial_i a = 
   J_i~,
\end{equation} 
and the source $J_i$ now is
\begin{equation} 
\label{J_i} 
\begin{split} 
  J_i &= - W \partial_i b^2  
  + 2 (\partial_i a)(\partial_r a) 
  + \e{-2A} (\Pi^j_i  c) (\partial_j b)
  + \frac12 (\partial_j b)(\partial_r e^j_i) 
  - \frac12 \e{-2A} \frac{\partial_j c}{\Box} \Box e^j_i\\ 
  &\quad   
  - \frac14 \partial_i \partial_r (e^j_k e^k_j)  
  + \frac12 e^j_k \partial_r \partial_j e^k_i 
  + \frac14 (\partial_i e^j_k)(\partial_r e^k_j) 
  - \frac12 \e{-2A} e^j_k \partial_j (\partial^k d_i -\partial_i 
d^k)\\ 
  &\quad 
  - \frac12 \e{-2A} (\partial_j e^k_i) (\partial^j d_k -\partial_k 
d^j) 
  - \frac12 \e{-2A} d_j \Box e^j_i 
  - \frac12 \e{-2A} (\partial_j b)(\partial^j d_i +\partial_i d^j)\\
  &\quad
  +2 (\partial_i \sigma)(\partial_r \sigma)~. 
\end{split} 
\end{equation} 

As in \cite{Bianchi:2003ug}, we first define $\ta=(W/W_\phi) a$ and
solve \eqref{eq_c} and \eqref{eq_bd} for $b$, $c$ and $d_i$, with the
results 
\begin{equation} 
\label{bcd_sol}  
\begin{split} 
  b &= -\frac{W_\phi^2}{W^2} \ta -\frac1{2W} \frac{\partial_i}{\Box} 
J^i~,\\ 
  \Box d_i &= -2 \e{2A} \Pi^j_i J_j~,\\ 
  \e{-2A} c &= \frac{W_\phi^2}{W^2} \partial_r \ta   
  - \frac1{4W} J_c +\frac{V}{W^2} \frac{\partial_i}{\Box} J^i~. 
\end{split} 
\end{equation} 
Then, we substitute \eqref{bcd_sol} into
\eqref{eq_a}, which yields the following second order ODE for $\ta$,  
\begin{equation} 
\label{eq_ta} 
  \left(D^2 +\e{-2A} \Box \right) \ta = 
  J_{\ta}~. 
\end{equation} 
We have abbreviated 
\begin{equation} 
\label{D2def} 
  D^2 = \left[ \partial_r + 2 \left(W_{\phi\phi} -\frac{W_\phi^2}{W} 
  -\frac{d}{d-1}W \right) \right] \partial_r~, 
\end{equation} 
and the source term in \eqref{eq_ta} is given by 
\begin{equation} 
\label{J_ta} 
  J_{\ta} = \frac{W}{W_\phi} J_a -\frac14 J_c -\frac12 
   \left[ \partial_r   
  + 2 \left( W_{\phi\phi} -\frac{W_\phi^2}{W} 
  -\frac{d}{d-1}W \right) \right] \frac{\partial_i}{\Box} J^i~. 
\end{equation} 
In the source \eqref{J_ta}, the first order terms of the solutions
\eqref{bcd_sol} should be substituted.

\subsection{The Equation for the Inert Scalar}
\label{inert_eq}
The equation for the inert scalar is quite easily obtained from
\eqref{eq_a} and \eqref{J_a} by dropping the terms with $W_\phi$ and
$V_\phi$ and generalizing the derivatives of $V$. One obtains
\begin{equation} 
\label{eq_sigma} 
  \left( \partial_r^2 -\frac{2d}{d-1} W \partial_r +\e{-2A} \Box
  -V_{\sigma\sigma} \right) \sigma = J_\sigma~, 
\end{equation} 
where the source $J_\sigma$ is 
\begin{equation} 
\label{J_s} 
\begin{split} 
  J_\sigma &=  \frac12 V_{\sigma\sigma\sigma} \sigma^2 
  + V_{\sigma\sigma\phi} \sigma a +
  2 V_{\sigma\sigma} \sigma b + (\partial_r \sigma)(\partial_r b) \\ 
  &\quad 
  +\e{-2A} \left[ -2 b\Box \sigma - (\partial^i b)(\partial_i \sigma)   
  + c \partial_r \sigma + 2\left( d^i +\frac{\partial^i c}{\Box} \right)  
  \partial_i \partial_r \sigma  \right.\\ 
  &\left.\quad   
  + \left( \partial_r  d^i + \partial_r \frac{\partial^i c}{\Box} 
  \right) \partial_i \sigma  
  - 2\frac{d-2}{d-1}W \left(d^i +\frac{\partial^i c}{\Box} \right) 
  \partial_i \sigma + e^i_j \partial_i \partial^j \sigma \right]~.
\end{split}  
\end{equation}  

\subsection{The Equation for the Graviton}
\label{grav_eq}
The tangential components of Einstein's equation provide the equations
of motion for the traceless transversal modes $e^i_j$.
Following the notation of \cite{Bianchi:2003ug}, we first write these
components in terms of time-slice hyper surface quantities, which gives  
\begin{multline}
\label{Einstein_ij}
  - \partial_r (n\K^i_j) + n^k \nabla_k(n\K^i_j) +(n\K^i_j) (n\K^k_k
  +\partial_r \ln n -n^k \partial_k \ln n )\\
  + n \nabla^i \partial_j n + (n\K^i_k) \nabla_j n^k 
  - (n\K^k_j) \nabla_k n^i + n^2 R^i_j +2 n^2 S^i_j =0~,
\end{multline}
where 
\begin{equation}
\label{S_ij}
  S^i_j = g^{ik} \left( \partial_k \phi \partial_j \phi + \partial_k
  \sigma \partial_j \sigma \right) + \frac{2}{d-1} \delta^i_j
  V(\phi,\sigma)~.
\end{equation}
In order to obtain an equation containing only the gauge invariants
$e^i_j$, $\ta$ and $\sigma$, we should expand \eqref{Einstein_ij} to
second order and substitute the solutions \eqref{bcd_sol} for $b$, $c$
and $d^i$. However, it is clear that the trace and the 
divergence of the resulting equation would both vanish by virtue of
the equation for $\ta$, \eqref{eq_ta},
because the scalar equation is implied by
Einstein's equation via the Bianchi identity. Thus, we can project
onto the traceless transversal components using the projector
\eqref{TTproj_def} in order to find the only independent equation that
is still missing. The result is 
\begin{equation}
\label{eq_e}
  \left( \partial_r^2 -\frac{2d}{d-1} W \partial_r +\e{-2A} \Box
  \right) e^i_j = J^i_j~,
\end{equation}
where the quadratic source terms are given by
\begin{equation}
\label{J_ij}
\begin{split}
  J^i_j &= \Pi^{ik}_{jl} \left\{ (\partial_r e^l_m) (\partial_r e^m_k) 
  -\e{-2A} \left[ e^m_n ( 2 \partial^l \partial_m e^n_k -\partial_m
  \partial^n e^l_k) 
  \phantom{\frac12} \right. 
  \phantom{\left(\left(\frac12\right)^2\right)} \right. \\
  &\quad \left.
  +\frac12 (\partial^l e^m_n)(\partial_k e^n_m) 
  +(\partial_m e^l_n)(\partial^n e^m_k) 
  -(\partial_m e^n_k)(\partial^m e^l_n) \right] \\
  &\quad 
  +2\left(\frac{W_\phi}{W}\right)^2 \left[ (\partial_m\partial_r
  e^l_k) \left(\frac{\partial^m}{\Box} \partial_r \ta \right) 
  + \e{-2A} \ta \Box e^l_k \right] 
  - \left[\partial_r \left(\frac{W_\phi}{W}\right)^2 \right] \ta
  \partial_r e^l_k \\
  &\quad 
  +2 \left(\frac{W_\phi}{W}\right)^4 \left[ 
  \left( \frac{\partial^l\partial_m}{\Box} \partial_r \ta \right) 
  \left( \frac{\partial^m\partial_k}{\Box} \partial_r \ta \right) 
  + \e{-2A} \partial^l \ta \partial_k \ta \right] \\
  &\quad \left. 
  + \left[\partial_r \left(\frac{W_\phi}{W}\right)^4 \right] \ta
  \frac{\partial^l\partial_k}{\Box} \partial_r \ta 
  -4\e{-2A} \left[  \left(\frac{W_\phi}{W}\right)^2 \partial^l\ta
  \partial_k \ta +\partial^l \sigma \partial_k \sigma \right]
  \right\}~.
\end{split}
\end{equation}

\section{Three-Point Functions}
\label{Sec_threepoint}

\subsection{General Form}
\label{general}
The dynamics of the bulk fields, which is governed by the equations
presented in the previous section, encodes the two- and three-point
functions of the dual operators. For completeness and in order to
outline our conventions for the presentation of the final results, we
shall in the following review how the correlations functions are
obtained from the (sub-leading) asymptotic behaviour of the bulk
fields. 

The gauge invariant fields $a$, $\sigma$ and $e^i_j$, which are the true
degrees of freedom of our bulk system, are dual to the operators
$\Op$, $\Sigma$ and $T^i_j$, respectively. Here and henceforth,
we denote by $T^i_j$ the \emph{traceless transversal
part} of the boundary energy-momentum tensor, 
\begin{equation}
\label{Tdef}
  T^i_j = \Pi^{ik}_{jl} \Theta^l_k~.
\end{equation}
The duality between the bulk fields and the boundary operators is made
explicit through the couplings 
\begin{equation}
\label{op_coupling}
  \int \rmd^d x\, \left[ \ha(x) \Op(x) 
  + \hat{\sigma}(x) \Sigma(x) + \frac12 \hat{e}^i_j(x) T^j_i(x) \right]~,
\end{equation}
where $\ha$, $\hat{\sigma}$ and $\hat{e}^i_j$ are the prescribed
asymptotic boundary values of the bulk fields $a$, $\sigma$ and
$e^i_j$, respectively. 

The other components of the energy momentum tensor are not
independent, because the (anomalous) 
Ward identities \cite{Bianchi:2001kw, Martelli:2002sp} imply the
operator identities 
\begin{equation}
\label{op_id}
  \partial_i \Theta^i_j =0~, \quad \Theta = \beta \Op~,
\end{equation}
which are valid in all correlation functions with distinct insertion
points. For vev flows we have $\beta=0$, whereas in operator flows
$\beta=-(d-\Delta)\hat{\bp}$, where $\Delta$ is the conformal
dimension of $\Op$ at the ultraviolet conformal fixed point of
the RG flow. 

In order to start, let us denote by $\psi$ a generic bulk field. 
Its behaviour in the asymptotically AdS region of the bulk space-time
is described by the generic expansion\footnote{In the general analysis, we
shall set the length scale of the asymptotic AdS region to one for
simplicity, \ie $L=1$.} 
\begin{equation}
\label{asympt}
  \psi(\rho,x) = \rho^{(d-\Delta)/2} [ \hpsi(x) + \cdots ] + 
  \rho^{\Delta/2} [ \rpsi(x) +\cdots ]~.
\end{equation}
Throughout this section, we shall use the variable $\rho=\e{-2r}$, so
that the asymptotic region is given by $\rho\to0$. Here, $\Delta$
denotes the conformal dimension of the operator dual to $\psi$,
which we shall call $\Psi$.\footnote{We restrict $\Delta$ by $d/2 <
  \Delta \le d$. The upper 
  bound means that we consider relevant or marginal deformations,
  whereas the lower bound stems from the standard AdS/CFT
  prescription. An extension down to the Breitenlohner-Freedman bound
  \cite{Breitenlohner:1982jf}, \ie for $d/2-1 < \Delta \le d/2$, has
  been developed in \cite{Klebanov:1999tb, Muck:1999kk}.} 
The ellipses in \eqref{asympt} 
stand for the sub-leading terms in the two series, which
are power series in $\rho$, whose coefficients depend locally on $\hpsi$
and $\rpsi$, respectively. If $\Delta-d/2$ is an integer, the leading
series contains also logarithms. The function $\rpsi$ is called the
\emph{response} function, and its non-trivial dependence on the
\emph{source} $\hpsi$ stems from the condition of regularity of $\psi$
in the bulk interior. Moreover, the response function $\rpsi$
determines the \emph{exact} one-point function of the operator
$\Psi$. More precisely, we have \cite{deHaro:2000xn,Bianchi:2001kw,
Martelli:2002sp} 
\begin{equation}
\label{gen:1pt}
  \vev{\Psi} = (2\Delta-d) \rpsi +\text{contact terms}~,
\end{equation}
where the contact terms are finite, but in principle scheme
dependent. We shall not be concerned with the contact terms in this
paper. A slightly different version of \eqref{gen:1pt} is
needed for the exact one-point function $\vev{T^i_j}$, which is 
\begin{equation}
\label{T_exact}
  \vev{T^i_j} = \frac14 (2\Delta-d) \check{e}^i_j 
  = \frac{d}4 \check{e}^i_j~. 
\end{equation}
Here, contact terms have been omitted.
The factor $1/4$ stems from the $1/4$ in front of the Einstein-Hilbert
term in the bulk action in our conventions \cite{Bianchi:2003ug}.

Let us focus our attention now on the response function, $\rpsi$. 
In order to determine it we first observe that $\psi$ satisfies a
second order ODE of the form    
\begin{equation} 
\label{eq_psi_form} 
  \left( \tilde{\nabla}^2 - M^2 \right) \psi = J_{\psi}~, 
\end{equation} 
where $M^2$ is an effective mass term, $\tilde{\nabla}$  
denotes the background covariant derivative, and $J_\psi$ is a higher
order source term. Thus, after defining a 
covariant Green's function by  
\begin{equation} 
\label{Green_def} 
  \left( \tilde{\nabla}^2 - M^2 \right) G(z,z') = 
  \frac{\delta(z-z')}{\sqrt{\tg(z)}}~, 
\end{equation} 
the general solution of \eqref{eq_psi_form} is given by  
\begin{equation} 
\label{psi_sol_form} 
  \psi(z) = \int \rmd^d y\, K(z,y) \hpsi(y) 
  + \int \rmd^{d+1} z'\,  
  \sqrt{\tg(z')} G(z,z') J_{\psi}(z')~. 
\end{equation} 
Here, $z$ is a short notation for the variables $(\rho,x)$, and $x$ 
and $y$ are boundary coordinates. Notice 
that the bulk integral is cut off at $\rho'=\varepsilon$, and 
also that $\rho\ge \varepsilon$, because of the regularization 
procedure. Moreover, $K(z,y)$ denotes the bulk-to-boundary
propagator of the field $\psi$. 
 
We are interested in the near-boundary behaviour of $\psi$, 
and it is very helpful that in asymptotically AdS spaces the Green's 
function asymptotically behaves as \cite{Muck:1999kk} 
\begin{equation} 
\label{Green_asympt} 
  G(z,z') \approx - \frac{\rho^{\Delta/2}}{2\Delta-d} K(x,z') 
+\cdots~. 
\end{equation}  
Hence, let us consider the field at the cut-off boundary by setting
$\rho=\epsilon$, and let us switch to momentum space, where we can use
momentum conservation of the propagators, $K(\rho,p;q)=K_p(\rho)
\delta(p+q)$. Then, \eqref{psi_sol_form} leads to
\begin{equation} 
\label{psi_asympt} 
  \psi(\varepsilon,p) \approx K_p(\varepsilon) \hpsi(p)  
  - \frac{\varepsilon^{\Delta/2}}{2\Delta-d} \int\limits_\varepsilon 
  \frac{\rmd \rho}{2\rho} \e{dA(\rho)}   
  K_p(\rho) J_{\psi}(\rho,p)~. 
\end{equation} 
From this general expression one must extract the response function
$\rpsi$, which represents, by virtue of \eqref{gen:1pt}, the exact
one-point function of the dual operator $\Psi$. The two-point
function $\vev{\Psi\Psi}$ can be read off easily from the asymptotic
behaviour of the bulk-to-boundary propagator $K_p$.

The various three-point functions that contain $\Psi$ are obtained
from the interaction integral in \eqref{psi_asympt}. 
In order to be more explicit, let us
consider a generic three-point function $\vev{\Psi_1\Psi_2\Psi_3}$,
where the $\Psi_n$, $n=1,2,3$, can be identical or different. Their
respective dependence on the momenta $p_n$ is implied. 
This three-point function can be obtained from the integral in
\eqref{psi_asympt} setting, \eg $\psi=\psi_1$. The term in
$J_{\psi_1}$ that is responsible for the three-point functions
$\vev{\Psi_1\Psi_2\Psi_3}$ has the generic form   
\begin{equation} 
\label{Jform} 
  J_{\psi_1}(\rho,p_1) = \int \rmd p_2\, \rmd p_3\,
  \delta(p_1+p_2+p_3) \X(p_1,-p_2,-p_3) K_2(\rho) K_3(\rho)
  \hpsi_2(-p_2) \hpsi_3(-p_3)~,
\end{equation} 
where we have substituted the linear solutions for the bulk fields
$\psi_2$ and $\psi_3$, and $\X$ is an operator containing also
derivatives with respect to $\rho$ (or, equivalently, $r$) acting on the
bulk-to-boundary propagators $K_2$ and $K_3$. It is also important to
notice the minus signs in front of $p_2$ and $p_3$. Differentiating
\eqref{Jform} with respect to the sources $\hpsi_2$ and $\hpsi_3$ and
substituting the result into \eqref{psi_asympt} yields the
three-point function of the form 
\begin{equation} 
\label{gen:corr} 
  \vev{\Psi_1 \Psi_2 \Psi_3} = 
  - \delta(p_1+p_2+p_3) 
  \int \rmd r\, \e{dA} \left(\frac{W_\phi}{W} \right)^2 
  \X_{123} K_1 K_2 K_3~,  
\end{equation} 
where $\X_{123}$ is again an operator acting on the bulk-to-boundary
propagators, which can be obtained easily from the source
$J_{\psi_1}$. 

Eqn.~\eqref{gen:corr} is the general expression for the three-point
functions, and our main work will consist in calculating the operators
$\X_{123}$. To explain this formula more precisely several
clarifications are in order. First, the  
factor $(W_\phi/W)^2$ has been inserted for continuity with
our previous calculations in \cite{Bianchi:2003ug}, where it
automatically appeared when using the field $\ta$ and its
corresponding bulk-to-boundary propagator, $\tK$. In fact, in the case
$\Psi_1=\Op$, the operator $\X_{123}$ can be read off directly
from the source $J_{\ta}$. In the other cases, $\Psi_1=\Sigma$ and
$\Psi_1=T^i_j$, we must multiply a factor $(W_\phi/W)^{-2}$ to
the operators read off from $J_\sigma$ and $J^i_j$,
respectively, in order to compensate for the factor in
\eqref{gen:corr}. Second, we must correctly carry out the functional
derivatives with respect to the sources, if $\Psi_2$ and $\Psi_3$ are
identical, which symmetrizes the operator of interest with respect to
the indices 2 and 3. Third, because of
\eqref{op_coupling} and \eqref{T_exact}, we must include a factor
$1/4$, if $\Psi_1 =T^i_j$, and a factor $2$ for each $T^i_j$ at the
positions $2$ and $3$. Forth, the reader is reminded of the minus
signs in front of $p_2$ and $p_3$ in \eqref{Jform}. Last, the integral 
in \eqref{gen:corr} is divergent in most cases. This is not much cause
of concern, because the divergences can be predicted and subtracted
using the counterterms of holographic renormalization. This is done
best on a case-by-case basis, and we shall illustrate it for one of
the three-point functions in the GPPZ flow in Sec.~\ref{divergences}.


\subsection{The Operators $\X_{123}$}
\label{3ptfuns}
We shall, in this subsection, calculate the operators $\X_{123}$ for all
ten independent three-point functions of $\Op$, $\Sigma$
and $T^i_j$. For this purpose, we take the equation of motion for
the dual field of one of the operators involved and consider the
source term containing the other two fields. If the source contains
terms with radial derivatives on both fields, we use a field
redefinition to eliminate those terms. Then, as explained in the
previous subsection, the operator $\X_{123}$ can be read off easily
from the new source. It will contain terms with radial derivatives
(with respect to $r$) acting on one of the bulk-to-boundary
propagators in the integral \eqref{gen:corr}. These derivatives are
denoted by $\partial_1$, $\partial_2$ and $\partial_3$, which act on 
$K_1$, $K_2$ and $K_3$, respectively. In most cases we need to
integrate by parts some of the terms in order to obtain a
Bose symmetric expression. The resulting boundary terms (as well as
those from the field redefinition) can be dropped, because we are not
concerned about the scheme dependent local terms of the three-point
functions. 

The three-point functions involving operators of different kinds can
obviously be calculated in more than one way. We have actually
performed all 18 possible calculations and cross-checked the results
for the mixed three-point functions. In the following, we shall
present the ten independent expressions starting with $\X_{OOO}$,
which we repeat from \cite{Bianchi:2003ug} for completeness.

\subsubsection*{The Correlator
$\vev{\Op\Op\Op}$}
This three-point function can be calculated only by considering those
terms of the source $J_{\ta}$ of \eqref{eq_ta} that are quadratic in
$\ta$. These are obtained by inserting the linear solutions for $b$
and $c$ into the general expression \eqref{J_ta} dropping $d_i$,
$e^i_j$ and $\sigma$. Thus, we find  
\begin{align} 
\notag  J_{\ta} &= \frac14 \left( \frac{W_\phi}{W} \right)^4  
  \left[ 2 \frac{\partial^i\partial_j}{\Box} 
  \left( \ta' \Pi^j_i \ta' \right) - (\Pi^j_i \ta') 
  \frac{\partial^i\partial_j}{\Box}  \ta' \right]  
  + 2 \left( \frac{W_\phi}{W} \right)^2  
  \frac{\partial^i \ta'}{\Box} \partial_i \ta'\\ 
\notag  &\quad +\frac12 \left[\partial_r \left( \frac{W_\phi}{W}
  \right)^4 \right]  
  \frac{\partial^i\partial_j}{\Box} \left( \ta \Pi^j_i \ta' 
  \right)  
  + \left[\partial_r \left( \frac{W_\phi}{W} \right)^2 \right]  
  \left[ \frac{\partial^i \ta'}{\Box}  \partial_i  \ta 
  - \frac{\partial^i}{\Box} \left( \ta' \partial_i \ta  
  \right) - \ta'  \ta \right] \\ 
\notag  &\quad -\frac12 \left( \frac{W_\phi}{W} \right)^4 \e{-2A} 
  \frac{\partial^i\partial_j}{\Box} \left(\ta \Box \Pi^j_i \ta 
  \right)  
  +  \frac12 \left[D^2 \left( \frac{W_{\phi\phi}}{W} - 
  \frac{W_\phi^2}{W^2} \right)\right] \ta^2\\
\label{OOO:J1} 
  &\quad + \left( \frac{W_\phi}{W} \right)^2 \e{-2A} \left[ 2\ta 
\Box \ta +  
  \frac{\partial^i}{\Box} \left[ (\partial_i \ta)(\Box \ta) \right] 
  -\frac12 (\partial^i \ta)(\partial_i \ta) \right]~,
\end{align} 
where $D^2$ is the second order differential operator defined in
\eqref{D2def}, and we have abbreviated $\ta'=\partial_r \ta$.

As in \cite{Bianchi:2003ug}, in order to facilitate the integrations
by parts, we remove the terms of $J_{\ta}$ with $r$-derivatives on
both fields by the field redefinition
\begin{equation} 
\label{OOO:redef} 
 \ta \to \ta +\frac18 \left( \frac{W_\phi}{W} \right)^4  
  \left[ 2 \frac{\partial^i\partial_j}{\Box} \left( \ta 
  \Pi^j_i \ta \right) - (\Pi^j_i \ta) 
  \frac{\partial^i\partial_j}{\Box} \ta \right]  
  + \left( \frac{W_\phi}{W} \right)^2  \frac{\partial^i \ta}{\Box} 
  \partial_i \ta~. 
\end{equation} 
In terms of this new field, the source becomes 
\begin{align} 
\notag 
  J_{\ta} &= \frac12 \left[\partial_r \left( \frac{W_\phi}{W}
  \right)^4 \right]  
  \left[ \frac{\partial^i\partial_j}{\Box} \left( \ta' 
         \frac{\partial^j\partial_i}{\Box} \ta \right) 
     -   \left(\frac{\partial^i\partial_j}{\Box} \ta'\right) 
         \left(\frac{\partial^j\partial_i}{\Box} \ta \right) 
\right]\\ 
\notag &\quad 
  - \left[ \partial_r \left( \frac{W_\phi}{W} \right)^2\right]  
  \left[ \frac{\partial^i \ta'}{\Box} \partial_i \ta 
  +2 \frac{\partial^i \ta}{\Box} \partial_i \ta'  
  + \frac{\partial^i}{\Box} \left( \ta' \partial_i \ta \right) 
  + \ta' \ta \right] \\ 
\notag &\quad  
  - \frac12 \left( \frac{W_\phi}{W} \right)^4 \e{-2A} 
  \left[ \ta \Box \ta - \frac{\partial^i\partial_j}{\Box} \left( \ta 
  \partial_i \partial^j \ta \right) 
  +\frac12 (\partial_i \ta)(\partial^i \ta) \right.\\ 
\notag &\quad \left.  
  - \frac{\partial^i\partial_j}{\Box} \left( (\partial_k \ta) 
  \frac{\partial_i \partial^j \partial^k}{\Box} \ta \right) 
  + \frac12 \left(\frac{\partial^i \partial^j \partial^k}{\Box} 
\ta\right)  
            \left(\frac{\partial_i \partial_j \partial_k}{\Box}
  \ta\right) \right] \\
\notag &\quad 
  + \left( \frac{W_\phi}{W} \right)^2 \e{-2A} 
  \left\{ 2\ta \Box \ta + \frac{\partial^i}{\Box} \left[ (\partial_i 
\ta) 
  (\Box \ta) \right] -\frac12 (\partial^i\ta)(\partial_i \ta) 
  - 2 \left(\frac{\partial^i\partial_j}{\Box} \ta\right)(\partial_i 
\partial^j \ta) 
  \right\} \\ 
\notag &\quad 
  +\frac12 \left[D^2 \left( \frac{W_{\phi\phi}}{W} - 
    \frac{W_\phi^2}{W^2} \right) \right] \ta^2  
  - \left[D^2 \left( \frac{W_\phi}{W} \right)^2\right]  
  \frac{\partial^i\ta}{\Box} \partial_i \ta \\ 
\label{OOO:J2} 
  &\quad -\frac18 \left[D^2 \left( \frac{W_\phi}{W} \right)^4\right]  
   \left[ \ta^2 -2 \frac{\partial^i\partial_j}{\Box} \left( \ta 
   \frac{\partial^j\partial_i}{\Box} \ta \right) + 
   \left(\frac{\partial^i\partial_j}{\Box} \ta \right)
   \left(\frac{\partial^j\partial_i}{\Box} \ta \right)\right]~. 
\end{align} 

We can read off $\X_{123}$ from \eqref{OOO:J2}, where we have to take
care to symmetrize the indices $2$ and $3$ because of the functional
derivation with respect to the sources. 
Using also momentum conservation, $p_1+p_2+p_3=0$, we find the operator
\begin{align} 
\notag 
  \X_{OOO} &= \frac12 \left[\partial_r \left(
  \frac{W_\phi}{W}\right)^4 \right]  
  \left[ \frac{(p_1\cdot p_2)^2}{p_1^2p_2^2} \partial_3 +  
         \frac{(p_1\cdot p_3)^2}{p_1^2p_3^2} \partial_2 +   
         \frac{(p_2\cdot p_3)^2}{p_2^2p_3^2} \partial_1  \right.\\ 
\notag &\quad \left. 
       - \frac{(p_2\cdot p_3)^2}{p_2^2p_3^2}  
         (\partial_1 +\partial_2 +\partial_3) \right] \\ 
\notag &\quad 
  + \left[\partial_r \left( \frac{W_\phi}{W}\right)^2 \right]  
  \left[  
  p_2\cdot p_3 \left(\frac1{p_2^2}+\frac1{p_3^2}\right) \partial_1 +  
  p_1\cdot p_3 \left(\frac1{p_1^2}+\frac1{p_3^2}\right) \partial_2
  \right.\\
\notag &\quad \left. 
  + p_1\cdot p_2 \left(\frac1{p_1^2}+\frac1{p_2^2}\right) \partial_3  
  - p_2\cdot p_3 \left(\frac1{p_2^2}+\frac1{p_3^2}\right)  
  (\partial_1 +\partial_2 +\partial_3) \right] \\ 
\notag &\quad 
  +\frac12 \left( \frac{W_\phi}{W}\right)^4 \e{-2A}  
  \left[ \frac12 (p_1^2+p_2^2+p_3^2) +  
   \frac{(p_1\cdot p_2)^3}{p_1^2p_2^2} +  
   \frac{(p_1\cdot p_3)^3}{p_1^2p_3^2} +   
   \frac{(p_2\cdot p_3)^3}{p_2^2p_3^2} \right] \\ 
\notag &\quad 
  + \left( \frac{W_\phi}{W}\right)^2 \e{-2A} 
  \left[ (p_2\cdot p_3)^2 \left(\frac1{p_2^2}+\frac1{p_3^2}\right) +  
         (p_1\cdot p_3)^2 \left(\frac1{p_1^2}+\frac1{p_3^2}\right)
  \right. \\
\notag & \quad \left. 
         + (p_1\cdot p_2)^2 \left(\frac1{p_1^2}+\frac1{p_2^2}\right) 
         - 2(p_1^2+p_2^2+p_3^2) \right] \\
\notag &\quad
  +\frac14 \left[D^2 \left( \frac{W_\phi}{W}\right)^4\right]  
  \left[ \frac{(p_1\cdot p_2)^2}{p_1^2p_2^2} +  
         \frac{(p_1\cdot p_3)^2}{p_1^2p_3^2} -   
         \frac{(p_2\cdot p_3)^2}{p_2^2p_3^2} \right]\\ 
\label{OOO:X1} 
  &\quad + \left\{ D^2 \left[ \frac{W_{\phi\phi}}{W} 
  -\left(\frac{W_\phi}{W}\right)^2  
  -\frac14 \left(\frac{W_\phi}{W}\right)^4 \right] \right\} 
  -\left[ D^2  \left(\frac{W_\phi}{W}\right)^2 \right]  
   p_2\cdot p_3 \left(\frac1{p_2^2}+\frac1{p_3^2}\right)~. 
\end{align} 
Because of the identity
\begin{equation}  
  (\partial_1 +\partial_2 +\partial_3) (\tK_1 \tK_2 \tK_3) = 
\partial_r (\tK_1 
  \tK_2 \tK_3)~, 
\end{equation} 
we can integrate these two terms of \eqref{OOO:X1} by parts in the 
integral in \eqref{gen:corr} taking note also of the identity 
\begin{equation} 
\label{D2ident} 
  \partial_r \left[\e{dA} \left(\frac{W_\phi}{W}\right)^2 \partial_r 
  \right] = \e{dA} \left(\frac{W_\phi}{W}\right)^2 D^2~. 
\end{equation} 
Hence, the last term in \eqref{OOO:X1} is cancelled, and the minus sign 
of the last term on the penultimate line is reversed, rendering the 
final result totally symmetric in the indices 1, 2 and 3, as it should
be. Thus, the final result is  
\begin{align} 
\notag 
  \X_{OOO} &= \frac12 \left[\partial_r \left(
  \frac{W_\phi}{W}\right)^4 \right] 
  \frac{(p_1\cdot p_2)^2}{p_1^2p_2^2} \partial_3   
  + \left[\partial_r \left( \frac{W_\phi}{W}\right)^2 \right] 
  p_1\cdot p_2 \left(\frac1{p_1^2}+\frac1{p_2^2}\right) \partial_3 \\ 
\notag &\quad  
  +\frac12 \left( \frac{W_\phi}{W}\right)^4 \e{-2A}  
  \left[ \frac12 p_1^2 + \frac{(p_1\cdot p_2)^3}{p_1^2p_2^2} \right]  
  + \left( \frac{W_\phi}{W}\right)^2 \e{-2A} 
  \left[ (p_1\cdot p_2)^2 \left(\frac1{p_1^2}+\frac1{p_2^2}\right) 
  -2p_1^2 \right] \\ 
\label{OOO:X2} &\quad  
  +\frac14 \left[D^2 \left( \frac{W_\phi}{W}\right)^4 \right]  
   \frac{(p_1\cdot p_2)^2}{p_1^2p_2^2}  
  +\frac13 \left\{D^2 \left[ \frac{W_{\phi\phi}}{W} 
  -\left(\frac{W_\phi}{W}\right)^2  
  -\frac14 \left(\frac{W_\phi}{W}\right)^4 \right] \right\}
  +\text{cyclic}~.  
\end{align}

\subsubsection*{The Correlator $\vev{\Op\Op T}$}
In order to calculate the correlator $\vev{\Op\Op T}$,
we consider the terms of $J_{\ta}$ that are of the form $\ta e^i_j$. 
These are 
\begin{equation}
\label{OOT:J1}
\begin{split}
  J_{\ta} &= \frac14 \left(\frac{W_\phi}{W}\right)^2 
     \left\{ 
       \frac{\partial_i \partial^j}{\Box} \left(\ta' {e^i_j}' \right) 
     - \left( \frac{\partial_i \partial^j}{\Box} \ta' \right) {e^i_j}'
     + \frac{\partial_i}{\Box} \left[ \left( \frac{\partial^j}{\Box}
     \ta' \right) \Box {e^i_j}' \right] \right\} \\
  &\quad
  + \frac14 \left[\partial_r \left(\frac{W_\phi}{W}\right)^2 \right] 
  \left\{ 
  2 \frac{\partial_i \partial^j}{\Box} \left(\ta {e^i_j}' \right) 
  + \frac{\partial_i}{\Box} \left[ \left( \frac{\partial^j}{\Box}
    \ta' \right) \Box e^i_j \right] \right\} \\
  &\quad 
  -\frac12 \left(\frac{W_\phi}{W}\right)^2  \e{-2A} 
   \frac{\partial_i \partial^j}{\Box} \left(\ta \Box e^i_j \right)
  + \e{-2A} e^i_j \partial_i \partial^j \ta~.
\end{split}
\end{equation}
The terms with two $r$-derivatives are removed by the field
redefinition
\begin{equation}
\label{OOT:redef}
  \ta \to \ta +\frac18 \left(\frac{W_\phi}{W}\right)^2 \left\{ 
       \frac{\partial_i \partial^j}{\Box} \left(\ta e^i_j \right) 
     - \left( \frac{\partial_i \partial^j}{\Box} \ta \right) e^i_j
     + \frac{\partial_i}{\Box} \left[ \left( \frac{\partial^j}{\Box}
     \ta \right) \Box e^i_j \right] \right\}~,
\end{equation}
which leads to the new source
\begin{equation}
\label{OOT:J2}
\begin{split}
  J_{\ta} &= \frac18 
   \left[ \partial_r \left(\frac{W_\phi}{W}\right)^2 \right]  
   \left\{ 
   \frac{\partial_i \partial^j}{\Box} \left(\ta {e^i_j}'
   \right) 
   +3 \left( \frac{\partial_i \partial^j}{\Box} \ta \right)
   {e^i_j}'  
   -3 \frac{\partial_i}{\Box} \left[ \left(
   \frac{\partial^j}{\Box} \ta \right) \Box {e^i_j}' \right] \right.\\
   &\quad \left.
   - 2\frac{\partial_i \partial^j}{\Box} \left(\ta' e^i_j \right) 
   + 2\left( \frac{\partial_i \partial^j}{\Box} \ta' \right) e^i_j 
  \right\} \\
  &\quad
  -\frac14 \left(\frac{W_\phi}{W}\right)^2  \e{-2A} \left\{ 
   \frac{\partial_i \partial^j}{\Box} \left( \partial_k \ta \partial^k
   e^i_j \right) 
   - \left( \frac{\partial_i \partial^j}{\Box} \partial_k \ta \right)
   \partial^k e^i_j 
   + \frac{\partial_i}{\Box} \left[ \left( 
   \frac{\partial^j \partial_k}{\Box} \ta \right) \partial^k \Box
   e^i_j \right] \right.\\
   &\quad \left.  
   +2 \frac{\partial_i \partial^j}{\Box} \left( \ta \Box e^i_j \right)
   \right\}\\
   &\quad 
   -\frac18 \left[ D^2 \left(\frac{W_\phi}{W}\right)^2 \right] 
   \left\{ 
     \frac{\partial_i \partial^j}{\Box} \left(\ta e^i_j \right) 
     - \left( \frac{\partial_i \partial^j}{\Box} \ta \right) e^i_j
     + \frac{\partial_i}{\Box} \left[ \left( \frac{\partial^j}{\Box}
     \ta \right) \Box e^i_j \right] \right\}
   + \e{-2A} e^i_j \partial_i \partial^j \ta~.
\end{split}
\end{equation}
From \eqref{OOT:J2} it is straightforward to read off the operator
$\X_{OOT}$, where we must remember to multiply by a factor of $2$ for
the $T$ insertion. Using momentum conservation, it can be
written in the form
\begin{equation}
\label{OOT:X1}
\begin{split}
  \X_{OOT}{}^i_j &= \frac14 \Pi_3{}^{ik}_{jl} \frac{p_{1k} p_2^l}{p_1^2 p_2^2}
   \left\{  
   8 \e{-2A} p_1^2 p_2^2 \phantom{\left(\frac12\right)^2} \right. \\
   &\quad
   +\left[\partial_r \left(\frac{W_\phi}{W}\right)^2 \right]
   \left[ (p_2^2 - p_1^2 + p_3^2) (\partial_1 +\partial_2
   +\partial_3) \right.\\
   &\quad \left.
   + (p_1^2 -p_2^2-p_3^2) \partial_1 + (p_2^2-p_1^2-p_3^2)\partial_2 
   + 2 (p_3^2 -p_1^2 -p_2^2)\partial_3 \right] \\
   &\quad
   + \left[ D^2 \left(\frac{W_\phi}{W}\right)^2 \right]
   (p_2^2 -p_1^2+p_3^2)  \\
   &\quad \left.
   + \left(\frac{W_\phi}{W}\right)^2 \e{-2A} 
   (p_1^4 +p_2^4+p_3^4 -2p_1^2 p_2^2 -2 p_1^2p_3^2 -2 p_2^2p_3^2 )
   \right\}~.
\end{split}
\end{equation}
Integrating the term containing $(\partial_1+\partial_2+\partial_3)$ by
parts we cancel the term on the penultimate line and end up with
\begin{equation}
\label{OOT:X2}
\begin{split}
  \X_{OOT}{}^i_j &= \frac14 \Pi_3{}^{ik}_{jl} 
   \frac{p_{1k} p_2^l}{p_1^2 p_2^2} 
   \left\{ 
   8 \e{-2A} p_1^2 p_2^2 \phantom{\left(\frac12\right)^2} \right. \\
   &\quad 
   +\left[\partial_r \left(\frac{W_\phi}{W}\right)^2 \right]
   \left[ (p_1^2 -p_2^2-p_3^2) \partial_1 
   + (p_2^2-p_1^2-p_3^2)\partial_2 
   + 2 (p_3^2 -p_1^2 -p_2^2)\partial_3 \right] \\
   &\quad \left.
   + \left(\frac{W_\phi}{W}\right)^2 \e{-2A} 
   (p_1^4 +p_2^4+p_3^4 -2p_1^2 p_2^2 -2 p_1^2p_3^2 -2 p_2^2p_3^2 )
   \right\}~.
\end{split}
\end{equation}
Obviously, the operator is symmetric in the indices 1 and 2, as it
should be.

\subsubsection*{The Correlator $\vev{\Op TT}$}
For this correlator we need to keep the terms in $J_{\ta}$ that are
quadratic in $e^i_j$. It is straightforward to find
\begin{equation}
\label{OTT:J1}
\begin{split}
  J_{\ta} &= \frac14 \Pi^i_j \left( {e^j_k}'{e^k_j}' \right) 
  +\frac14 \left(W_{\phi\phi} -\frac{W_\phi^2}{W} \right) 
  \left[ 2 \Pi^i_j (e^j_k {e^k_i}') 
  -\frac{\partial^i}{\Box} (\partial_i e^j_k {e^k_j}' ) \right]\\
  &\quad 
  +\frac14 \e{-2A} \left\{ \frac{\partial^i\partial_j}{\Box}
  \left( e^j_k \Box e^k_i \right) 
  +\frac34 (\partial_i e^j_k)(\partial^i e^k_j)
  -\frac12 (\partial_i e^j_k)(\partial^k e^i_j)
  +\frac12 \frac{\partial^i}{\Box} \left[ (\partial_i e^j_k) \Box
  e^k_j \right] \right\}~.
\end{split}
\end{equation}
The first term is removed by the field redefinition
\begin{equation}
\label{OTT:redef}
  \ta \to \ta +\frac18 \Pi^i_j (e^j_k e^k_i)~,
\end{equation}
which leads to the new source
\begin{equation}
\label{OTT:J2}
\begin{split}
  J_{\ta} &= -\frac14 \left(W_{\phi\phi} -\frac{W_\phi^2}{W} \right) 
  \frac{\partial^i}{\Box} (\partial_i e^j_k {e^k_j}') \\
  &\quad 
  +\frac18 \e{-2A} \left\{ 
  -\frac12 (\partial_i e^j_k)(\partial^i e^k_j)
  + \frac{\partial^i}{\Box} \left[ (\partial_i e^j_k) \Box
  e^k_j \right] \right\}~.
\end{split}
\end{equation}
From \eqref{OTT:J2} we easily read off the operator (remember the
factor $4$ for the two $T$ insertions)
\begin{equation}
\label{OTT:X}
\begin{split}
  \X_{OTT}{}^{ik}_{jl} &= 
  \frac1{p_1^2} \Pi_2{}^{im}_{jn} \Pi_3{}^{kn}_{lm} 
  \left[ \left(W_{\phi\phi} -\frac{W_\phi^2}{W} \right) 
  ( p_1 \cdot p_2 \partial_3 + p_1\cdot p_3 \partial_2 ) \right.\\
  &\quad \left.
  +\frac14 \e{-2A} (p_1^4 +p_2^4 +p_3^4 -2 p_1^2 p_2^2 -2 p_1^2 p_3^2
  -2 p_2^2 p_3^2) \right]~.
\end{split}
\end{equation}
This operator is symmetric in the indices 2 and 3 because of
the functional derivation with respect to the sources $\hat{e}^i_j$.

\subsubsection*{The Correlator $\vev{\Op\Op\Sigma}$}
Due to \eqref{VIphi}, there is no term of the form $\ta\sigma$ in the
source $J_{\ta}$. Hence, this correlation function vanishes,
\begin{equation}
\label{OOS}
  \vev{\Op\Op\Sigma} =0~.
\end{equation}

\subsubsection*{The Correlator $\vev{\Op\Sigma\Sigma}$}
The terms of $J_{\ta}$ that are quadratic in $\sigma$ are
\begin{equation}
\label{OSS:J}
\begin{split}
  J_{\ta} &= -2 \left( W_{\phi\phi} -\frac{W_\phi^2}{W} \right)
  \frac{\partial^i}{\Box} \left(\sigma' \partial_i \sigma \right) +
  \frac12 \left( \frac{W}{W_\phi} V_{\phi\sigma\sigma} -2
  V_{\sigma\sigma} \right) \sigma^2 \\
  &\quad 
  +\e{-2A} \left\{ \frac{\partial^i}{\Box} \left[(\partial_i
  \sigma)(\Box \sigma) \right] -\frac12 (\partial^i
  \sigma)(\partial_i\sigma) \right\}~.
\end{split}
\end{equation}
As terms with two $r$-derivatives are absent, we can directly read off
the operator $\X_{O\Sigma\Sigma}$,
\begin{equation}
\label{OSS:X}
\begin{split}
  \X_{O\Sigma\Sigma} &= 
  \frac{W}{W_\phi} V_{\phi\sigma\sigma} -2 V_{\sigma\sigma}
  +\frac2{p_1^2} \left[ \left( W_{\phi\phi} -\frac{W_\phi^2}{W}
  \right) (p_1 \cdot p_2 \partial_3 + p_1 \cdot p_3 \partial_2 )
  \right.\\
  &\quad \left. 
  +\frac14 \e{-2A} (p_1^4 +p_2^4 +p_3^4 -2 p_1^2 p_2^2 -2 p_1^2 p_3^2
  -2 p_2^2 p_3^2) \right]~.
\end{split}
\end{equation}

\subsubsection*{The Correlator $\vev{\Op\Sigma T}$}
As there is no term of the form $e^i_j\sigma$ in the
source $J_{\ta}$, this correlation function vanishes,
\begin{equation}
\label{OST}
  \vev{\Op \Sigma T^i_j} =0~.
\end{equation}

\subsubsection*{The Correlator $\vev{\Sigma\Sigma\Sigma}$}
This three-point function can only be obtained by considering those
terms of the source $J_\sigma$ of \eqref{eq_sigma} that are quadratic
in $\sigma$. There is only one such term, which is 
\begin{equation}
\label{SSS:J}
  J_\sigma = \frac12 V_{\sigma\sigma\sigma} \sigma^2~,
\end{equation}
from which we directly read off
\begin{equation}
\label{SSS:X}
  \X_{\Sigma\Sigma\Sigma} =  
   \left(\frac{W_\phi}{W}\right)^{-2} V_{\sigma\sigma\sigma}~.
\end{equation}
Remember that the factor $(W_\phi/W)^{-2}$ is needed for the
general convention \eqref{gen:corr}.

\subsubsection*{The Correlator $\vev{\Sigma\Sigma T}$}
For this three-point function we consider the term in $J_\sigma$ that
is bi-linear in $e^i_j$ and $\sigma$, which is
\begin{equation}
\label{SST:J}
  J_\sigma = \e{-2A} e^i_j \partial_i \partial^j \sigma~,
\end{equation}
from which we can easily read off the operator (remember the factor
$2$ for the $T$ insertion)
\begin{equation}
\label{SST:X}
  \X_{\Sigma\Sigma T}{}^i_j = 2
  \left(\frac{W_\phi}{W}\right)^{-2} \e{-2A}
  \Pi_3{}^{ik}_{jl} p_{1k} p_2^l~.
\end{equation}

\subsubsection*{The Correlator $\vev{\Sigma T T}$}
As there is no $(e^i_j)^2$ term in $J_\sigma$, this correlator
vanishes,
\begin{equation}
\label{STT}
  \vev{\Sigma T^i_j T^k_l} =0~.
\end{equation}

\subsubsection*{The Correlator $\vev{TTT}$}
This three-point function is obtained from those terms of the source
$J^i_j$ of \eqref{eq_e} that are quadratic in $e^i_j$. These are 
\begin{equation}
\label{TTT:J1}
\begin{split}
  J^i_j &= \Pi^{ik}_{jl} \left\{ (e^l_m)' (e^m_k)' 
  -\e{-2A} \left[ 
  e^m_n ( 2 \partial^l \partial_m e^n_k -\partial_m \partial^n e^l_k)
  \phantom{\frac12} \right.\right.\\
  &\quad \left. \left.
  +\frac12 (\partial^l e^m_n)(\partial_k e^n_m) 
  + (\partial_m e_n^l)(\partial^n e^m_k) 
  - (\partial_m e_n^l)(\partial^m e^n_k) \right] \right\}~.
\end{split}
\end{equation}
After the field redefinition 
\begin{equation}
\label{TTT:redef}
  e^i_j \to e^i_j + \frac12 \Pi^{ik}_{jl} e^l_m e^m_k 
\end{equation}
we obtain the new source
\begin{equation}
\label{TTT:J2}
\begin{split}
  J^i_j &= - \Pi^{ik}_{jl} \e{-2A} \left[ 
  e^m_n ( 2 \partial^l \partial_m e^n_k -\partial_m \partial^n e^l_k)
  +\frac12 (\partial^l e^m_n)(\partial_k e^n_m) 
  + (\partial_m e_n^l)(\partial^n e^m_k) \right]~.
\end{split}
\end{equation}
Remembering to include the factors $(W_\phi/W)^{-2}$ for our
convention, $1/4$ for the first $T$ and $4=2\cdot 2$ for the other two
$T$s, the operator $\X_{TTT}$ is easily found from \eqref{TTT:J2}, 
\begin{equation}
\label{TTT:X}
\begin{split}
  \X_{TTT}{}^{ikm}_{jln} &= 
   \e{-2A} \left(\frac{W_\phi}{W} \right)^{-2} 
   \Pi_1{}^{ii'}_{jj'} \Pi_2{}^{kk'}_{ll'}\Pi_3{}^{mm'}_{nn'} \\
   &\quad \times 
   \left[ 2 \left( 
              p_{1k'} p_2^{j'} \delta^{n'}_{i'} \delta^{l'}_{m'} 
            + p_{1m'} p_3^{j'} \delta^{n'}_{k'} \delta^{l'}_{i'} 
            + p_{2m'} p_3^{l'} \delta^{j'}_{k'} \delta^{n'}_{i'} 
         \right) \right. \\
  &\quad \left.
  +  p_{1m'} p_2^{n'} \delta^{j'}_{k'} \delta^{l'}_{i'} 
  +  p_{1k'} p_3^{l'} \delta^{n'}_{i'} \delta^{j'}_{m'} 
  +  p_{2i'} p_3^{j'} \delta^{n'}_{k'} \delta^{l'}_{m'} \right]~.
\end{split}
\end{equation}
Obviously, this operator is Bose symmetric.

\section{GPPZ Flow Correlation Functions}
\label{GPPZ_corrs}

In this section, we shall apply our general results to the GPPZ flow
with the aim of calculating the glueball scattering amplitudes.
To begin, we shall repeat in subsection~\ref{two_pts}
the calculation of the bulk-to-boundary propagators, which also encode
the two-point functions, as they will be needed for the explicit
calculation of scattering amplitudes later. The mass spectra and decay
constants of the associated glueball states are read off in a standard
fashion. Then, by isolating the on-shell poles of the bulk-to-boundary
propagators the external legs of the Feynman diagrams are explicitly
amputated. 
In subsection~\ref{GPPZ:threept}, we provide the expressions for the
non-zero three-point functions in the GPPZ flow. These automatically 
encode the three-particle scattering amplitudes by amputating the
external legs. It is a fortunate fact that the on-shell (amputated)
bulk-to-boundary propagators are polynomials. Thus, the radial
integral in the scattering amplitudes is elementary.
Throughout this section, we use the radial variable $u$ defined by
$u=1-\rho=1-\e{-2r}$, as is customary for the GPPZ flow. The
asymptotic AdS region of the bulk space-time is given by $u\to 1$.

\subsection{Bulk-to-Boundary Propagators and Two-Point Functions}
\label{two_pts}

\subsubsection*{Active Scalar}
Let us begin with the bulk-to-boundary propagator for the active
scalar. The linear equation of motion \eqref{eq_ta} for $\ta$ becomes
(in momentum space) 
\begin{equation}
\label{twopt_eq_ta}
  \left[ u(1-u) \partial_u^2 +(2-2u) \partial_u -\frac{p^2}{4} \right]
  \ta = 0~.
\end{equation}
This is a hypergeometric equation, whose solution, which
is regular for $u=0$, is readily found. We have to be somewhat careful
with the normalization, because the bulk-to-boundary propagator for
$a$, $K=(W_\phi/W) \tK$, should satisfy the generic asymptotic
expansion with a unit coefficient of the leading term. 
Hence, we find the bulk-to-boundary propagator $\tK$ 
\begin{equation}
\label{twopt_a_sol}
  \tK_p (u) = \frac{\sqrt{3}}{2} 
    \Gamma \left(\frac{3+\alpha}2 \right) 
    \Gamma \left(\frac{3-\alpha}2 \right) 
  \F\left( \frac{1+\alpha}2, \frac{1-\alpha}2; 2; u \right)~,
\end{equation}
with $\alpha= \sqrt{1-p^2}$. Its asymptotic behaviour is
\cite{Abramowitz}
\begin{equation}
\label{twopt_a_asym}
  \tK_p (u) \approx  \frac{\sqrt{3}}{2} 
  \left[ 1 + \frac{p^2}4 (1-u) \ln(1-u) 
  + \frac12 H^O(p) (1-u) +\cdots \right]~.
\end{equation}
The function $H^O(p)$, which is related to the two-point function by
\begin{equation}
\label{twopt_a_2pt}
  \vev{\mathcal{O}(p)\mathcal{O}(q)}= \delta(p+q) H^O(p)~,
\end{equation} 
is given by
\begin{equation}
\label{twopt_a_H}
  H^O(p) = \frac{p^2}2 \left[ 
     \psi \left( \frac{3+\alpha}2 \right) 
   + \psi \left( \frac{3-\alpha}2 \right) -\psi(2) -\psi(1) \right]~.
\end{equation}
The spectrum of poles becomes clear after
rewriting $H^O$ in a series representation using the formula
\cite{Gradshteyn}
\begin{equation}
\label{psi_series}
  \psi(x) -\psi(y) = \sum\limits_{k=0}^{\infty} 
  \left( \frac1{y+k} -\frac1{x+k} \right)~.
\end{equation}
This yields
\begin{equation}
\label{twopt_a_Hser}
  H^O(p) = \frac{p^4}2 \sum\limits_{k=1}^\infty 
  \frac{2k+1}{k(k+1)[4k(k+1)+p^2]}~.
\end{equation}
Thus, we find particles with the masses
\begin{equation}
\label{twopt_a_spectrum}
  m_O^2= 4k(k+1)~, \quad k=1,2,3,\ldots~.
\end{equation} 
The residues at the poles, which represent the decay
constants \cite{Peskin}, are 
\begin{equation}
\label{twopt_a_res}
  |f^O_k|^2 = 8k(k+1)(2k+1)~.
\end{equation}

For an on-shell momentum, $p^2=-4k(k+1)$, we have $\alpha=2k+1$, and
the hypergeometric function in \eqref{twopt_a_sol} truncates to a
polynomial, whereas one of the $\Gamma$-functions in the normalization
factor has a pole. As in \cite{Bianchi:2003ug} we identically
re-write \eqref{twopt_a_sol} as 
\begin{equation}
\label{twopt_a_sol2}
  \tK_p (u) =  \frac{\sqrt{3}}{2} 
    \Gamma \left(\frac{1+\alpha}2 \right) 
    \Gamma \left(\frac{1-\alpha}2 \right)
    (1-u) \frac{\rmd}{\rmd u}  
  \F\left( \frac{1+\alpha}2, \frac{1-\alpha}2; 1; u \right)~,
\end{equation}  
and amputate the external leg as
\begin{equation}
\label{twopt_a_amput}
  \tK_p (u) = \frac{|f^O_k|}{p^2 +4k(k+1)} 
    \left[  \sqrt{\frac{3(2k+1)(k+1)}{2k}} 
    (1-u) \LeP_{k-1}^{(1,1)}(2u-1) \right] 
    + \text{regular}~,
\end{equation}
where $\LeP_k^{(\alpha,\beta)}$ denotes a Jacobi polynomial of degree
$k$ \cite{Gradshteyn}, 
\begin{equation} 
\label{JacPol} 
  \LeP_k^{(\alpha,\beta)}(2u-1) = \sum_{n=0}^{k} 
  \binom{k+\alpha}{n} \binom{k+\beta}{k-n} u^n (u-1)^{k-n}~.
\end{equation}

\subsubsection*{Inert Scalar}
Next, let us consider the inert scalar. 
The linear equation of motion \eqref{eq_sigma} for $\sigma$ becomes
\begin{equation}
\label{twopt_eq_sigma}
  \left[ u(1-u) \partial_u^2 +(2-2u) \partial_u -\frac{p^2}{4} + 2\right]
  \frac{\sigma}{\sqrt{1-u}} = 0~.
\end{equation}
Again, we have a hypergeometric equation, whose solution, regular for
$u=0$ and normalized properly, is
\begin{equation}
\label{twopt_s_sol}
  K^\sigma_p (u) = 
    \Gamma \left(\frac{3+\beta}2 \right) 
    \Gamma \left(\frac{3-\beta}2 \right) \sqrt{1-u}\,
  \F\left( \frac{1+\beta}2, \frac{1-\beta}2; 2; u \right)~,
\end{equation}
with $\beta= \sqrt{9-p^2}$. Its asymptotic behaviour is very similar
to \eqref{twopt_a_asym},
\begin{equation}
\label{twopt_s_asym}
  K^\sigma_p (u) \approx \sqrt{1-u} 
  \left[ 1 + \frac{p^2-8}4 (1-u) \ln(1-u) 
  + \frac12 H^\Sigma(p) (1-u) +\cdots \right]~,
\end{equation}
where $H^\Sigma(p)$, which embodies the two-point function
$\vev{\Sigma\Sigma}$, is given by 
\begin{equation}
\label{twopt_s_H}
  H^\Sigma(p) = \frac{p^2-8}2 \left[ 
     \psi \left( \frac{3+\beta}2 \right) 
   + \psi \left( \frac{3-\beta}2 \right) -\psi(2) -\psi(1) \right]~.
\end{equation}
Using \eqref{psi_series}, we re-write \eqref{twopt_s_H} as 
\begin{equation}
\label{twopt_s_Hser}
  H^\Sigma(p) = \frac{(p^2-8)^2}2 \sum\limits_{k=1}^\infty 
  \frac{2k+1}{k(k+1)[4(k-1)(k+2)+p^2]}~.
\end{equation}
Thus, we obtain the mass spectrum 
\begin{equation}
\label{twopt_s_spectrum}
  m_\Sigma^2= 4(k-1)(k+2)~, \quad k=1,2,3,\ldots~,
\end{equation}
and the respective residues are 
\begin{equation}
\label{twopt_s_res}
  |f^\Sigma_k|^2 = 8k(k+1)(2k+1)~.
\end{equation}
Notice that the mass spectrum \eqref{twopt_s_spectrum} includes a
massless particle. 

On-shell, we have $\beta=2k+1$, and, as for the active scalar,
the hypergeometric function in \eqref{twopt_s_sol} truncates to a
polynomial, whereas one of the $\Gamma$-functions in the normalization
factor has a pole. Hence, after first rewriting \eqref{twopt_s_sol} as 
\begin{equation}
\label{twopt_s_sol2}
  K^\sigma_p (u) = 
    \Gamma \left(\frac{1+\beta}2 \right) 
    \Gamma \left(\frac{1-\beta}2 \right)
    (1-u)^{3/2} \frac{\rmd}{\rmd u}  
  \F\left( \frac{1+\beta}2, \frac{1-\beta}2; 1; u \right)~,
\end{equation}  
we amputate the external leg as
\begin{equation}
\label{twopt_s_amput}
  K^\sigma_p (u) = \frac{|f^\Sigma_k|}{p^2 +4(k-1)(k+2)} 
    \left[ \sqrt{\frac{2(2k+1)(k+1)}{k}} 
    (1-u)^{3/2} \LeP_{k-1}^{(1,1)}(2u-1) \right] 
    + \text{regular}~.
\end{equation}

\subsubsection*{Graviton}
Last, we consider the traceless transversal part of the graviton,
$e^i_j$. Its linear equation of motion \eqref{eq_e} becomes
(in momentum space) 
\begin{equation}
\label{twopt_eq_e}
  \left[ u(1-u) \partial_u^2 +(2-u) \partial_u -\frac{p^2}{4} \right]
  e^i_j = 0~.
\end{equation}
Also this is a hypergeometric equation, and its regular and properly
normalized solution is 
\begin{equation}
\label{twopt_e_sol}
  K^e_p (u) = 
    \Gamma \left(\frac{4+\gamma}2 \right) 
    \Gamma \left(\frac{4-\gamma}2 \right) 
  \F\left( \frac{\gamma}2, -\frac{\gamma}2; 2; u \right)~,
\end{equation}
with $\gamma= \sqrt{-p^2}$. Its asymptotic behaviour is
\cite{Abramowitz}
\begin{equation}
\label{twopt_e_asym}
  K^e_p (u) \approx 
  1 - \frac{p^2}4 (1-u) - \frac{p^2(p^2+4)}{32}(1-u)^2 \ln(1-u) 
  + \frac12 H^T(p) (1-u)^2 +\cdots~.
\end{equation}
The function $H^T(p)$, which embodies the two-point function by
\begin{equation}
\label{twopt_e_2pt}
  \vev{T^i_j(p)T^k_l(q)}= \Pi^{ik}_{jl} \delta(p+q) H^T(p)~,
\end{equation} 
is given by
\begin{equation}
\label{twopt_e_H}
  H^T(p) = - \frac{p^2(p^2+4)}{16} \left[ 
     \psi \left( \frac{4+\gamma}2 \right) 
   + \psi \left( \frac{4-\gamma}2 \right) -\psi(3) -\psi(1) \right]~.
\end{equation}
After re-writing \eqref{twopt_e_H} as a series using
\eqref{psi_series}, we obtain
\begin{equation}
\label{twopt_e_Hser}
  H^T(p) = -\frac{p^2(p^2+4)^2}{8} \sum\limits_{k=1}^\infty 
  \frac{k+1}{k(k+2)[4(k+1)^2+p^2]}~.
\end{equation}
Thus, we find particles with the masses 
\begin{equation}
\label{twopt_e_spectrum}
  m_T^2= 4(k+1)^2~,\quad k=1,2,3,\ldots~,
\end{equation}
and the respective decay rates are
\begin{equation}
\label{twopt_e_res}
  |f^T_k|^2 = 8k(k+2)(k+1)^3~.
\end{equation}
These states describe spin-two glueballs, but we postpone a discussion
of the spin structure to Sec.~\ref{scatt}, where it is needed for the
calculation of the scattering amplitudes.

On shell, we have $\gamma=2(k+1)$, and
the hypergeometric function in \eqref{twopt_e_sol} truncates to a
polynomial, whereas one of the $\Gamma$-functions in the normalization
factor has again a pole. As before, we first re-write \eqref{twopt_e_sol} as 
\begin{equation}
\label{twopt_e_sol2}
  K^e_p (u) = 
    \Gamma \left(\frac{2+\gamma}2 \right) 
    \Gamma \left(\frac{2-\gamma}2 \right) 
    (1-u)^2 \frac{\rmd}{\rmd u}
  \F\left( \frac{2+\gamma}2, \frac{2-\gamma}2; 1; u \right)~,
\end{equation}
and then amputate the external leg as
\begin{equation}
\label{twopt_e_amput}
  K^e_p (u) = \frac{|f^T_k|}{p^2 +4(k+1)^2} 
    \left[ 2 \sqrt{\frac{2(k+1)(k+2)}{k}} 
    (1-u)^2 \LeP^{(2,1)}_{k-1}(2u-1) \right] 
    + \text{regular}~.
\end{equation}


\subsection{Three-point Functions and Scattering Amplitudes}
\label{GPPZ:threept}
In this subsection, we shall re-write our general results for the
three-point functions for the case of the GPPZ flow. 
In order to fix convention, we shall write the three-point functions
in the form
\begin{equation}
\label{threept:corr}
  \vev{\Psi_1\Psi_2\Psi_3} = -\delta(p_1+p_2+p_3) \int \rmd u\,
  \Y_{123} K_1 K_2 K_3~,
\end{equation}
where the operators $\Y_{123}$ are related to the operators
$\X_{123}$, which were presented in section~\ref{3ptfuns}, by
\begin{equation}
\label{threept:Ydef}
  \Y_{123} = 
  \frac{\e{4A(u)}}{2(1-u)} \left(\frac{W_\phi}{W}\right)^2 \X_{123} 
  = \frac23 \left(\frac{u}{1-u}\right)^2 \X_{123}~.
\end{equation}
The symbols $\partial_n$, $n=1,2,3$, that appear in the expressions
for $\Y_{123}$ are 
now derivatives with respect to $u$ acting on the respective $K_n$.
The integral in \eqref{threept:corr} is divergent in all cases but
one. The degree of divergence will be indicated, but the discussion of
the divergences and how they are removed is postponed to subsection
\ref{divergences}.  

It follows from \eqref{threept:corr} that the irreducible scattering
amplitudes take the form
\begin{equation}
\label{threept:M}
  \mathcal{M}_{123} = - \int \rmd u\,
  \Y_{123} \hat{K}_1 \hat{K}_2 \hat{K}_3~,
\end{equation}
where $\hat{K}_n$, $n=1,2,3$, denote the amputated bulk-to-boundary
propagators, which are given by the expressions in the brackets in
\eqref{twopt_a_amput}, \eqref{twopt_s_amput} and
\eqref{twopt_e_amput}. Since the amputated bulk-to-boundary
propagators are polynomials in $u$, the integral in \eqref{threept:M}
is in general elementary. 

For the GPPZ flow, there are only six independent non-zero three-point 
functions involving $\mathcal{O}$, $\Sigma$ and $T^i_j$. In addition to 
$\vev{\mathcal{O}\mathcal{O}\Sigma}$, $\vev{\mathcal{O}T\Sigma}$ and
$\vev{TT\Sigma}$, which vanish in general, also $\vev{\Sigma\Sigma\Sigma}$ 
vanishes in the GPPZ background, because of $V_{\sigma\sigma\sigma}=0$. 
The operators $\Y_{123}$ for the six non-trivial three-point functions are 
listed below. 

The operator $\Y_{OOO}$ was calculated already in
\cite{Bianchi:2003ug}, but we shall repeat it here for completeness.  
From \eqref{OOO:X2} and \eqref{threept:Ydef} we obtain
\begin{equation}
\label{threept:OOO} 
\begin{split}
  \Y_{OOO} &= \frac{32}{27} \left\{ 
  -4 u^2 (1-u)  
  \frac{(p_1\cdot p_2)^2}{p_1^2p_2^2} \partial_3   
  - 3 u^2  
  p_1\cdot p_2 \left(\frac1{p_1^2}+\frac1{p_2^2}\right) \partial_3 \right. \\ 
  &\quad  
  +\frac12 u(1-u)   
  \left[ \frac14 (p_1^2+p_2^2) 
  + \frac{(p_1\cdot p_2)^3}{p_1^2p_2^2} \right]  
  + \frac34 u  
  \left[ (p_1\cdot p_2)^2 \left(\frac1{p_1^2}+\frac1{p_2^2}\right) 
  -(p_1^2+p_2^2) \right] \\ 
  &\quad \left. 
  + 2 u(3u-2)  
   \frac{(p_1\cdot p_2)^2}{p_1^2p_2^2}  
  + u \left[2(1-u) +\frac13 \right] \right\}
  +\text{cyclic}~. 
\end{split}
\end{equation}
Since $\tK\sim\sqrt{3}/2$ to leading order, the integral
\eqref{threept:corr} for $\vev{\mathcal{O}\mathcal{O}\mathcal{O}}$ is
finite. 

For the correlator $\vev{\mathcal{O}\mathcal{O}T}$ we find 
\begin{equation}
\label{threept:OOT}
\begin{split}
   \Y_{OOT}{}^i_j &=  \Pi_3{}^{ik}_{jl} 
   p_{1k} p_2^l \frac2{9p_1^2 p_2^2} 
   \left\{ 
   6 \frac{u}{1-u} p_1^2 p_2^2 \right. \\ 
   & \quad
   - 4 u^2
   \left[ (p_1^2 -p_2^2-p_3^2) \partial_1 
   + (p_2^2-p_1^2-p_3^2)\partial_2 
   + 2 (p_3^2 -p_1^2 -p_2^2)\partial_3 \right] \\
   & \left. \phantom{\frac12}
   + u
   (p_1^4 +p_2^4+p_3^4 -2p_1^2 p_2^2 -2 p_1^2p_3^2 -2 p_2^2p_3^2 )
   \right\}~.
   \end{split}
\end{equation}
The first term in the braces leads to a logarithmic divergence in the
integral in \eqref{threept:corr}. 

For the three-point function $\vev{\mathcal{O}TT}$ we obtain
\begin{equation}
\label{threept:OTT}
\begin{split}
  \Y_{OTT}{}^{ik}_{jl} &= 
  \Pi_2{}^{im}_{jn} \Pi_3{}^{kn}_{lm} \frac1{6p_1^2}  \frac{u}{1-u} 
  \left\{ -4u 
  [ (p_3^2 - p_1^2 - p_2^2) \partial_3 + (p_2^2 - p_1^2 - p_3^2)
  \partial_2 ] \right.\\
  &\quad \left.
  + (p_1^4 +p_2^4 +p_3^4 -2 p_1^2 p_2^2 - 2 p_1^2 p_3^2 -2 p_2^2
  p_3^2) \right\}~. 
\end{split}
\end{equation}
Again, the first term in the bracket leads to a logarithmic
divergence in \eqref{threept:corr}.

Next, for $\vev{\mathcal{O}\Sigma\Sigma}$ we find 
\begin{equation}
\label{threept:OSS}
\begin{split}
  \Y{}_{O\Sigma\Sigma} &= 
  \frac{4u(2-3u)}{(1-u)^2}
  +\frac1{3 p_1^2} \frac{u}{1-u} \left\{ -4u [(p_3^2 - p_1^2 - p_2^2)
  \partial_3 + (p_2^2 - p_1^2 - p_3^2) \partial_2 ] \right. \\
  &\quad \left. 
  + (p_1^4 +p_2^4 +p_3^4 -2 p_1^2 p_2^2 -2
   p_1^2 p_3^2 -2 p_2^2 p_3^2) \right\}.
\end{split}
\end{equation}
The first term leads to a logarithmic divergence in the integral in
\eqref{threept:corr}, because of the leading behaviour
$K^\sigma\sim\sqrt{1-u}$. 

The expression for $\vev{\Sigma\Sigma T}$ is rather simple,
\begin{equation}
\label{threept:SST}
  \Y_{\Sigma\Sigma T}{}^i_j =
  \Pi_3{}^{ik}_{jl} p_{1k} p_2^l \frac{u}{(1-u)^2}~.
\end{equation}
Again, the integral in \eqref{threept:corr} is logarithmically divergent.

Finally, the operator for the correlator $\vev{TTT}$ takes the form
\begin{equation}
\label{threept:TTT}
\begin{split}
  \Y_{TTT}{}^{ikm}_{jln} &= 
   \Pi_1{}^{ii'}_{jj'} \Pi_2{}^{kk'}_{ll'}\Pi_3{}^{mm'}_{nn'}
   \frac{1}{2}\frac{u}{(1-u)^2} \\
   &\quad \times 
   \left[ 2 \left( 
              p_{1k'} p_2^{j'} \delta^{n'}_{i'} \delta^{l'}_{m'} 
            + p_{1m'} p_3^{j'} \delta^{n'}_{k'} \delta^{l'}_{i'} 
            + p_{2m'} p_3^{l'} \delta^{j'}_{k'} \delta^{n'}_{i'} 
         \right) \right. \\
  &\quad \left.
  +  p_{1m'} p_2^{n'} \delta^{j'}_{k'} \delta^{l'}_{i'} 
  +  p_{1k'} p_3^{l'} \delta^{n'}_{i'} \delta^{j'}_{m'} 
  +  p_{2i'} p_3^{j'} \delta^{n'}_{k'} \delta^{l'}_{m'} \right]~.
\end{split}
\end{equation}
This operator leads to a linear divergence in the integral in
\eqref{threept:corr}.


\subsection{Finding Divergences from Holographic Renormalization}
\label{divergences}
As anticipated in the general discussion of the formula
\eqref{gen:corr} and seen explicitly in the previous subsection, 
the integral that formally embodies the three-point
functions may be divergent. These divergences are understood
and predicted by the results of holographic renormalization. Hence, it
is possible to subtract the divergences and to obtain finite results.
In this subsection, we shall illustrate this procedure using the
integral for the three-point function $\vev{\mathcal{O}\Sigma\Sigma}$.  

Before starting, we would like to emphasize also that the divergences
are irrelevant for the physical three-particle scattering
amplitudes. In fact, the integral \eqref{threept:M} involving the
amputated bulk-to-boundary propagators are finite. This is easily
understood by looking, \eg at the bulk-to-boundary 
propagator $\tK$ in the form \eqref{twopt_a_sol2} and its amputated
form  \eqref{twopt_a_amput}. We have a factor $(1-u)$ in both
formulae, which cancels a factor $(1-u)^{-1}$ in the integral
measure. However, the hypergeometric function in \eqref{twopt_a_sol2}
goes like $\ln(1-u)$ for $u\to1$ for generic $\alpha$, 
whereas it is a polynomial on-shell. This means that, if the integral
is divergent for a generic three-point function
because of negative powers of $(1-u)$ in the integrand, in its
amputated on-shell version we have a sufficient number of $(1-u)$
factors to cancel them.

Let us start our illustration by summarizing the predictions of holographic
renormalization for the divergences. We shall use the Hamilton-Jacobi
method \cite{Martelli:2002sp}, which is particularly simple to apply. 
Let us remind the reader that in the Hamilton-Jacobi method one first
calculates the counterterms by solving a set of algebraic equations,
and then one continues to obtain the exact one-point functions 
\eqref{gen:1pt}. 

Thus, in order to find the counterterms, we first expand the potential
$V(\phi,\sigma)$ about the fixed point $\phi=\sigma=0$. From
\eqref{GPPZ_V} we find 
\begin{equation}
\label{V_exp}
  V(\phi,\sigma) = -3 -\frac32 \left(\phi^2 + \sigma^2\right) 
      -\frac13 \left(\phi^4 -3 \sigma^4 +6 \phi^2 \sigma^2\right)
  +\cdots~.
\end{equation}
After applying the procedure of solving the descent equations as
explained in \cite{Martelli:2002sp} we arrive at the following
counterterms,\footnote{The same counterterms were obtained in
  \cite{Bianchi:2003bd} using the standard method of holographic
  renormalization.} 
\begin{equation}
\label{counterterms}
\begin{split}
  S_{\text{c.t.}} &= \int \rmd^4 x\, \sqrt{g} \left\{ \frac32
  +\frac12\left(\phi^2 +\sigma^2\right) -\frac18 R 
  - \ln \varepsilon \left[ \frac23\left(\phi^2 \sigma^2-\sigma^4\right)
  \right. \right. \\
  &\quad \left. \left. 
  -\frac14 \left(\nabla_i \phi \nabla^i \phi + \nabla_i \sigma \nabla^i
  \sigma\right) +\frac1{24} \left(\phi^2 +\sigma^2\right)R  
  + \frac1{32} \left( R^i_j R^j_i -\frac13 R^2 \right) \right] 
  \right\}~.  
\end{split}
\end{equation}

In order to find the exact one-point functions one writes down a
series expansion for the fields. In the case of the field $\phi$ we
have 
\begin{equation}
\label{phi_exp}
  \phi(x,\varepsilon) = \hat{\phi}(x) \varepsilon^{1/2} + \tilde{\phi}(x) 
  \varepsilon^{3/2}\ln \varepsilon + \check{\phi}(x) \varepsilon^{3/2}
  +\cdots~. 
\end{equation}
It is now clear that the logarithmic divergences of the three-point
function integrals are understood as contributions to the logarithmic
term $\tilde{\phi}$ stemming from the interactions. Furthermore, it is
possible to make an exact statement using holographic renormalization,
because the term $\tilde{\phi}$ is determined uniquely by the fact
that the exact one-point function $\vev{\mathcal{O}}$ is finite. For the
counterterms \eqref{counterterms} this yields \cite{Martelli:2002sp}
\begin{equation} 
\label{tildephi}
  \tilde{\phi} = -\frac23 \hat{\phi} \hat{\sigma}^2 -\frac14
  \hat{\nabla}^2 \hp -\frac1{24} \hp \hat{R}~.
\end{equation}
Focussing on $\vev{\mathcal{O}\Sigma\Sigma}$, \eqref{phi_exp} and
\eqref{tildephi} imply that the second term in \eqref{psi_asympt}
contains a term  
\[ -\frac23 \hat{\phi} \hat{\sigma}^2 \varepsilon^{3/2} \ln
\varepsilon~, \]   
from which in turn follows that 
the integral in \eqref{gen:corr} with the operator
$\X_{O\Sigma\Sigma}$ contains the logarithmic divergence
\begin{equation}
\label{OSS_div}
  \frac83 \hat{\bar{\phi}} \ln \varepsilon = \frac8{\sqrt{3}} \ln
  \varepsilon~. 
\end{equation}

Similarly, it becomes also clear from the counterterms
\eqref{counterterms} that  
the integral for $\vev{\mathcal{O}\mathcal{O}\mathcal{O}}$ is finite,
the one for $\vev{TTT}$ is linearly divergent (the $R$ counter term
contains $e^i_j$ to all orders), and the others that are
not zero diverge logarithmically. These are precisely the divergences
observed explicitly in subsection \ref{GPPZ:threept}.

We shall now confirm the prediction \eqref{OSS_div} by calculating the
divergence directly. It stems from the terms on the first line of
\eqref{threept:OSS}. Substituting them into the integral
\eqref{threept:corr}  
and considering only the leading behaviour of the bulk-to-boundary
propagators \eqref{twopt_a_asym} and \eqref{twopt_s_asym} yields 
\begin{equation}
  \int\limits^{1-\varepsilon} 
  \rmd u\, \frac{\sqrt{3}}2 \left[ 
   \frac{4u(2-3u)}{1-u} - \frac83 \frac{u^2}{1-u}  
   \frac{p_1 \cdot (p_2+p_3)}{p_1^2}  
   \sqrt{1-u} \partial_u \sqrt{1-u} \right]~.
\end{equation}
By virtue of momentum conservation this gives to leading order
\begin{equation}
  \int\limits^{1-\varepsilon} 
  \rmd u\, \left(- \frac8{\sqrt{3}} \right) \frac1{1-u} =
  \frac8{\sqrt{3}} \ln \varepsilon~,
\end{equation}
which confirms \eqref{OSS_div}. 

Having understood the origin of the divergence, one can proceed to
remove it. This is done by subtracting the term from the integral that
contributes to $\tilde{\phi}$, \ie by adding 
\[  \frac{8}{\sqrt{3}} \frac1{1-u} \]
to the integrand.  One could repeat the above
discussion for all cases, but we shall be content with
this illustration.


\section{Glueball Scattering Amplitudes and Discussion}
\label{scatt}
At this point we have at our disposal the necessary ingredients to
calculate the scattering amplitudes for various three-particle
processes of interest. In order to restore the proper physical
dimensions, we should re-introduce the asymptotic AdS length scale $L$,
defined by $L^4=4\pi g_s N {\alpha'}^2 = \lambda_t {\alpha'}^2$, where
$\lambda_t= g_{YM}^2 N$ is the 't~Hooft coupling. This is simply done
by replacing $p$ by $pL$ everywhere, so that $\Op(p)\to \Op(pL)=
\Op(p)/L$. ($\Op(p)$ has length dimension $-1$ corresponding to
$\Op(x)$ having dimension $3$. The same relation holds for $\Sigma(p)$,
while $T^i_j(p)$ has dimension $0$.) Furthermore, the
results for the correlation functions should be multiplied by a
numerical factor $[N^2/(2\pi^2)]\times (2\pi)^4$, where the factor
$N^2/(2\pi^2)$ takes into account the 5-dimensional Newton
constant, while the $(2\pi)^4$ stems from our convention for the
$\delta$-function in momentum space. We shall not do this explicitly,
but only note that, after canonically normalizing the operators at the
UV fixed point, which absorbs the factor $N^2$ in the two-point
functions, the three-point functions are suppressed by a factor $1/N$. 

Before starting, however, we shall briefly summarize the particle
spectrum found in subsection~\ref{two_pts} in order to facilitate the
ensuing discussion.

The operator $\Op$, which is part of the $\mathcal{N}=1$ chiral
anomaly multiplet, $\mathcal{A} = \tr (\Phi^{2})$, creates the glueball
states $\Op_k$ of spin zero, which have masses
\begin{equation}
\label{Aspec}
(mL)^2 = 4 k(k+1) \qquad k=1,2,\ldots~.
\end{equation} 
The operator $\Sigma$ is part of the ``Lagrangian'' multiplet
$\mathcal{S} = \tr (W^2 + \cdots)$ and creates the glueball states
$\Sigma_k$ of spin zero, which have masses
\begin{equation}
\label{Sspec}
(mL)^2 = 4 (k-1)(k+2) \qquad k=1,2,\ldots~.
\end{equation}
Finally, the operator $T^i_j$, which is part of the $\mathcal{N}=1$
supercurrent multiplet  $\mathcal{J}_{\alpha \dot{\alpha}} = \tr
(W_{\alpha} \bar{W}_{\dot{\alpha}}+ \cdots )$, creates the spin-two
glueballs $T_k$ with the masses 
\begin{equation}
\label{Jspec}
(mL)^2 = 4 (k+1)^{2} \qquad k=1,2, \ldots~.
\end{equation}

The spin structure of the spin-two states $T_k$ can be described as
follows. In spin space, the operator $T^i_j$ is a symmetric, traceless
and transversal $4\times 4$ matrix, which forms a spin-two
representation of the ``little group'' for massive particles, 
\ie the subgroup of those Lorentz transformations that leave a given
massive momentum four-vector $p^i$ invariant. In fact, it is
instructive to choose as a basis a set of five matrices
$\varepsilon_r(p)^i_j$, with $r=1,2,3,4,5$, satisfying the
orthogonality relations 
\begin{equation}
\label{eps_ortho}
  \varepsilon_r(p)^i_j \varepsilon_s(p)^j_i = \delta_{rs}~,
\end{equation}
in addition to the transversality and tracelessness conditions
\begin{equation}
\label{eps_tt}
  \varepsilon_r(p)^i_j p_i= 0 =\varepsilon_r(p)^i_i~.
\end{equation}
Using such a basis, the traceless transversal projector can be written
in the form 
\begin{equation}
\label{TT_proj_eps}
  \Pi^{ik}_{jl} = \sum\limits_{r=1}^5 \varepsilon_r(p)^i_j
  \varepsilon_r(p)^k_l~.
\end{equation}
Eqn.~\eqref{TT_proj_eps} can also be regarded as a completeness
relation. 
The polarization matrices take a particularly simple form in the rest
frame, where they are given by
\begin{equation}
\label{eps_restframe}
  \varepsilon_r = \begin{pmatrix} 0         & \mathbf{0}  \\
                                  \mathbf{0}& \hat{\varepsilon}_r
                  \end{pmatrix}~,
\end{equation}
where the $\hat{\varepsilon}_r$ form a basis of symmetric and
traceless $3\times3$ matrices. 

The spectrum of states includes also the superpartners of the states
listed above, which have, obviously, the same masses. The spectrum of
low-lying states is shown in Fig.~\ref{plot_spectra}. 
\FIGURE{\includegraphics{GPPZ_masses}%
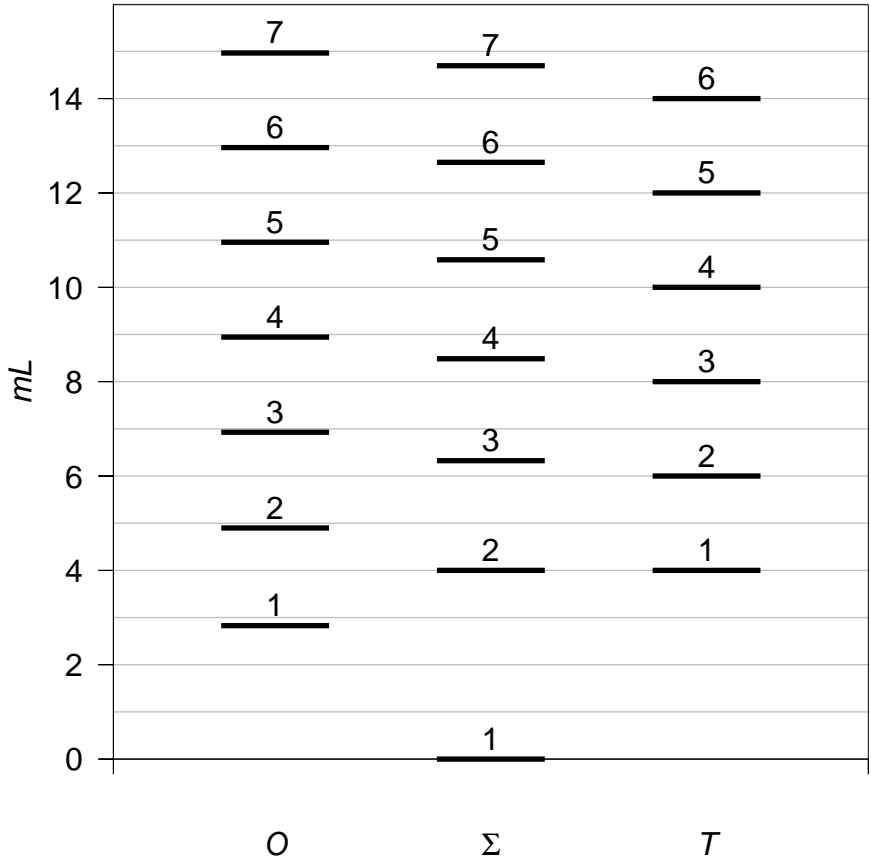
\caption{Spectra of states with $mL<15$. \label{plot_spectra}}}

We shall turn now to the scattering amplitudes, which, in an
equivalent fashion, describe the decay of a glueball into two other
glueballs, or the creation of that glueball in a two-glueball
collision. These processes are constrained by phase space such that
the mass of the decaying (or the created) glueball must be at least as
large as the sum of the masses of the decay products (or the colliding
particles). It will turn out that the amplitudes for most processes,
which are allowed by phase space, vanish.

In order to simplify the discussion, we have chosen to consider the
unpolarized amplitudes for processes involving the spin-two glueballs
$T_k$. This is achieved by first projecting the generic amplitude
containing indices $i$ and $j$ for each external spin-two state onto a
particular polarization using the polarization matrices
$\varepsilon_r(p)^i_j$, then taking the square and finally summing
over all polarizations $r$. For the amplitudes of $T\Op\Op$ processes
this implies 
\begin{align}
\notag
  |\mathcal{M}_{TOO}|^2 &= \sum\limits_{r=1}^5
   \left[\varepsilon_r(p_1)^i_j \mathcal{M}_{TOO}{}^j_i\right]
   \left[\varepsilon_r(p_1)^k_l \mathcal{M}_{TOO}{}^l_k\right]\\
\label{TOOunpol_1}
  &= \mathcal{M}_{TOO}{}^j_i \Pi_1{}^{ik}_{jl}
  \mathcal{M}_{TOO}{}^l_k~,
\end{align}
where we have used the completeness relation \eqref{TT_proj_eps}.
As $\mathcal{M}_{TOO}{}^i_j$ is of the form 
\[ 
 \mathcal{M}_{TOO}{}^i_j = \left(\Pi_1{}^{ik}_{jl}p_{2k}
 p_3^l\right) \tilde{\mathcal{M}}_{TOO}~,
\]
we find from \eqref{TOOunpol_1}
\begin{align}
\notag
  |\mathcal{M}_{TOO}|^2 &= 
  \left(p_{2i} p_{2k} p_3^j p_3^l \Pi_1{}^{ik}_{jl} \right)
   |\tilde{\mathcal{M}}_{TOO}|^2\\
\label{TOOunpol_2}
  &= \frac1{24p_1^4} \left( p_1^4 + p_2^4 + p_3^4 - 2p_1^2 p_2^2
  -2p_1^2p_3^2 -2p_2^2p_3^2\right)^2 |\tilde{\mathcal{M}}_{TOO}|^2~.
\end{align}
The last line is easily established in the rest frame of the $T$
particle. The unpolarized amplitude $|\mathcal{M}_{T\Sigma\Sigma}|^2$
is obtained in the same fashion.

Similarly, we might define the unpolarized amplitudes
$|\mathcal{M}_{OTT}|^2$ and $|\mathcal{M}_{TTT}|^2$, which give rise
to more cumbersome pre-factors involving the external momenta. We
shall not provide their explicit expressions, because our numerical
analysis will indicate that the amplitudes of all processes of these
kinds, which are allowed by phase space, vanish. Hence, the overall
factor is of no importance. 

The actual calculation of the scattering amplitudes \eqref{threept:M}
can be easily implemented on a computer.\footnote{We have used
MAXIMA. A script is available from the authors.} The numerical results
for the unpolarized glueball decay amplitudes for the glueball states
with $k\le10$ are listed in appendix~\ref{decay_amps}. Here, we
summarize our findings and the resulting decay channels. 

The states $\Op_1$ and $\Sigma_1$ turn out to be stable glueballs. For
$\Sigma_1$ this is natural, because it is massless, but for $\Op_1$ it
results from the fact that the only allowed process, $\Op_1\to \Sigma_1
+\Sigma_1$, has a zero amplitude. In contrast, the process  $T_1\to
\Sigma_1 +\Sigma_1$ is allowed and occurs. 

A glueball $\Op_k$ with $k>1$ decays mainly into two $\Sigma$
glueballs. It can decay into $\Sigma_k+\Sigma_1$, and into
$\Sigma_i+\Sigma_j$ such that $i+j=k$, but the latter processes are
severely restricted by phase space. In fact, for $k\ge10$, only the
processes $\Op_k\to\Sigma_{k-2}+\Sigma_2$ and
$\Op_k\to\Sigma_{k-1}+\Sigma_1$ are allowed. Furthermore, $\Op_k$ can
decay into $\Op_i +T_j$ such that $i+j=k-1$, although these decay
channels are much less probable. All other allowed processes have zero
amplitudes. In particular, there are no decays of the form $\Op\to
T+T$ and $\Op\to \Op+\Op$. 

The decay of a glueball $\Sigma_k$ with $k>1$ must contain exactly one
$\Sigma$ glueball amongst the products. The main decay channels are
into $\Sigma_i +T_j$ such that $i+j=k-1$, and into $\Op_{k-1}
+\Sigma_1$. The decay into $T_{k-1} +\Sigma_1$ also occurs, but has a
much smaller probability. Again, all other allowed processes have
vanishing amplitude. 

Finally, the glueballs $T_k$ decay mainly into two $\Sigma$ glueballs,
the main channels being into $\Sigma_i+\Sigma_j$ such that $i+j=k$,
and into $\Sigma_k+\Sigma_1$. Decays into $\Op_i+\Op_j$ such that
$i+j=k$ are also possible, but less probable. As before, all other
allowed decay processes have vanishing amplitude. 

Although our conclusions about the vanishing amplitudes of many decay
channels stem from the numerical analysis of the decay amplitudes up
to $k=20$, we believe that there is a deeper reason, which is to be
sought in some orthogonality relation of the Jacobi polynomials. We
shall not try to give a more rigorous proof of these statements.

A comparison of our results with similar data from lattice simulations of
$\mathcal{N}=1$ SYM theory, when they become available, would be very
interesting.

The three-point function $\vev{\Op\Sigma\Sigma}$ has been calculated also
by Bianchi and Marchetti \cite{Bianchi:2003bd}. Although their formula
differs from our \eqref{threept:OSS}, it is possible to show,
using integrations by parts and the equations of motion for the
bulk-to-boundary propagators, that the two bulk integrals differ only by
boundary terms. In the three-point function, these boundary terms
would constitute contact terms, which we should drop, because none of
us has done a reliable analysis of the contact terms. In the amplitudes, the
boundary terms vanish because of factors of $u$ and $(1-u)$. In fact,
our numerical amplitudes agree completetly.

In conclusion, the holographic analysis of the three-point functions
for the operators $\Op$, $\Sigma$ and $T^i_j$ has yielded precise
predictions for the glueball scattering amplitudes in the GPPZ
flow. As the GPPZ flow shares some features with pure $\mathcal{N}=1$
SYM theory (in particular confinement), 
it potentially sheds light on the IR dynamics of the
latter. However, one should be cautious to draw too quick a
conclusion. The GPPZ flow is a particular, unstable, case of a
two-parameter family of $\mathcal{N}=1$ holographic RG flow
backgrounds describing the mass deformation of $\mathcal{N}=4$ SYM
theory \cite{Girardello:1999bd}. In a generic member of this family 
both scalars considered in this paper, $\phi$ and $\sigma$, are
active, and the latter describes a gaugino condensate, which is a
necessary ingredient in the vacuum structure of $\mathcal{N}=1$ SYM
theory. However, all of these backgrounds are singluar, and it is
unclear how to choose amongst the parameters the right values that
describe an $\mathcal{N}=1$ vacuum. By analogy with other gravity
duals of $\mathcal{N}=1$ SYM theory (\eg the Maldacena-Nu\~nez
solution \cite{Maldacena:2000yy}), one might argue that a truely
10-dimensional mechanism---unknown at present---will resolve the bulk
singularities, thereby fixing the parameters and isolating the vacuum. 

Hence, one should regard the GPPZ flow as a toy model, whose
qualitative features exist also in the theory with the true
vacuum. These features include the existence of glueball states and
the preferred glueball decay channels, although the numerics of the
glueball masses and scattering amplitudes will be affected by the
non-zero gluino condensate and the singularity resolution. It might,
of course, have been better to consider a generic background of the
two-parameter family of solutions in order to describe at least the effect
of the gluino condensate. Unfortunately, there are two technical
difficulties already at the linearized level making this problem much
harder to tackle. First, the two active scalars present in these
backgrounds couple to each other through the potential leading to a
fourth order differential equation. Second, although the
traceless transversal components of the metric decouple from all other
fields, their equation of motion is not analytically solvable. Further
progress on these issues is, therefore, very desirable.

\acknowledgments
We would like to thank Massimo Bianchi for his collaboration in the
early stage of this project and for helpful discussions, as well as
Kostas Skenderis for sharing his experience in holographic
renormalization with us. 

This work was supported in part by INFN, by MIUR (contract
2003-023852), by NATO (contract PST.CLG.978785) and by the European
Community's Human Potential Programme (contracts HPRN-CT-2000-00122,
HPRN-CT-2000-00148 and HPRN-CT-2000-00131).

\begin{appendix}

\section{Useful Relations for the GPPZ Flow}
\label{GPPZ_rels}
We summarize here a number of relations for the GPPZ background. For
simplicity, we set the asymptotically AdS length scale to unity, \ie
$L=1$. 

The potential $V(\phi,\sigma)$ that gives rise to the GPPZ flow with
$\phi$ as an active scalar was found in \cite{Girardello:1999bd}. It
is given in terms of a superpotential $W(\phi,\sigma)$ by 
\begin{equation}
\label{Vdef}
  V(\phi,\sigma) = 
  \frac12 \left(\frac{\partial W}{\partial\phi}\right)^2 
  +\frac12  \left(\frac{\partial W}{\partial\sigma}\right)^2 
  - \frac43 W^2~,
\end{equation}
where 
\begin{equation}
\label{GPPZ_W}
  W(\phi,\sigma) = - \frac34 \left[ 
   \cosh \left(\frac{2\phi}{\sqrt{3}}\right) 
   +\cosh (2\sigma) \right]~.
\end{equation}
Hence, we have
\begin{equation}
\label{GPPZ_V}
  V(\phi,\sigma) = - \frac38 \left[ 
   \cosh^2 \left(\frac{2\phi}{\sqrt{3}} \right)  
   + 4 \cosh \left(\frac{2\phi}{\sqrt{3}} \right)\cosh (2\sigma) 
   - \cosh^2 (2\sigma) +4 \right]~.
\end{equation}

The GPPZ background solution is
\begin{equation}
\label{GPPZ_background}
  \e{2\bp/\sqrt{3}} = \frac{1+\e{-r}}{1-\e{-r}}~,\qquad
  \bar{\sigma} =0~,\qquad
  \e{2A} = \e{2r} - 1~.
\end{equation}
From \eqref{GPPZ_background} we easily find the background source,
\begin{equation}
\label{GPPZ_b_source}
  \hat{\bar{\phi}} = \sqrt{3}~.
\end{equation}

For the GPPZ background (for $\bar{\sigma}=0$), there are a number of
identities that simplify the calculations with the potentials and its
derivatives, namely 
\begin{equation}
\label{GPPZ_pot_ident}
\begin{aligned}
  W_{\phi\phi}           &= \frac43 W + 1~,\quad 
  &\frac{W_\phi^2}{W}    &= \frac43 W +2~,\quad 
  & V                    &= -W\left( \frac23 W-1 \right)~, \\
  V_{\sigma\sigma}       &= 8W +9~, 
  & V_{\sigma\sigma\phi} &= 8 W_\phi~,
 &V_{\sigma\sigma\sigma} &=0~.
\end{aligned}
\end{equation}

Finally, it is useful to introduce the variable
\begin{equation}
\label{GPPZ_u_def}
  u = 1 - \e{-2r}~,
\end{equation}
in terms of which the following relations hold,
\begin{equation}
\label{GPPZ_W_rels}
\begin{aligned}
  \frac{du}{dr} &= 2(1-u)~, \quad          
  & \e{-2A}     &= \frac{1-u}{u}~,\\
  W             &= -\frac{3}{2u}~,    
  & W_\phi      &= -\sqrt{3} \frac{\sqrt{1-u}}{u}~.
\end{aligned}
\end{equation}

\vfill
\pagebreak[4]

\section{Numerical Results for the Glueball Decay Amplitudes}
\label{decay_amps}
In this appendix we provide lists of all non-zero scattering
amplitudes ($|\mathcal{M}|^2$) 
for the decays of the glueballs $\Op_k$, $\Sigma_k$ and
$T_k$ with $k\le10$. Only the amplitudes for decay processes allowed
by the phase space are given. The amplitudes involving $T$ glueballs 
are the unpolarized ones as defined in Sec.~\ref{scatt}. 

\begin{center}
\begin{tabular}{|c|r||c|r||c|r|} 
\hline 
 & & & & $T_{1}\to \Sigma_{1}+\Sigma_1$ & $ 184.32 $ \\ 
\hline 
 $T_{2}\to \Sigma_{1}+\Sigma_{1}$  & $ 103.68 $ & 
 $T_{2}\to    \Op_{1}+   \Op_{1}$  & $ 0.08 $ & 
$T_{2}\to \Sigma_{2}+\Sigma_1$ & $ 81.633 $ \\ 
\hline 
 $T_{3}\to \Sigma_{2}+\Sigma_{1}$  & $ 105.8 $ & 
 $T_{3}\to    \Op_{2}+   \Op_{1}$  & $ 0.145 $ & 
$T_{3}\to \Sigma_{3}+\Sigma_1$ & $ 51.429 $ \\ 
\hline 
 $T_{4}\to \Sigma_{3}+\Sigma_{1}$  & $ 102.86 $ & 
 $T_{4}\to    \Op_{3}+   \Op_{1}$  & $ 0.179 $ & 
$T_{4}\to \Sigma_{4}+\Sigma_1$ & $ 38.099 $ \\ 
 $T_{4}\to \Sigma_{2}+\Sigma_{2}$  & $ 117.55 $ & 
 $T_{4}\to    \Op_{2}+   \Op_{2}$  & $ 0.287 $ & 
 & \\ 
\hline 
 $T_{5}\to \Sigma_{4}+\Sigma_{1}$  & $ 99.967 $ & 
 $T_{5}\to    \Op_{4}+   \Op_{1}$  & $ 0.197 $ & 
$T_{5}\to \Sigma_{5}+\Sigma_1$ & $ 30.847 $ \\ 
 $T_{5}\to \Sigma_{3}+\Sigma_{2}$  & $ 119.01 $ & 
 $T_{5}\to    \Op_{3}+   \Op_{2}$  & $ 0.367 $ & 
 & \\ 
\hline 
 $T_{6}\to \Sigma_{5}+\Sigma_{1}$  & $ 97.601 $ & 
 $T_{6}\to    \Op_{5}+   \Op_{1}$  & $ 0.209 $ & 
$T_{6}\to \Sigma_{6}+\Sigma_1$ & $ 26.368 $ \\ 
 $T_{6}\to \Sigma_{4}+\Sigma_{2}$  & $ 118.3 $ & 
 $T_{6}\to    \Op_{4}+   \Op_{2}$  & $ 0.415 $ & 
 & \\ 
 $T_{6}\to \Sigma_{3}+\Sigma_{3}$  & $ 123.23 $ & 
 $T_{6}\to    \Op_{3}+   \Op_{3}$  & $ 0.481 $ & 
 & \\ 
\hline 
 $T_{7}\to \Sigma_{6}+\Sigma_{1}$  & $ 95.705 $ & 
 $T_{7}\to    \Op_{6}+   \Op_{1}$  & $ 0.217 $ & 
$T_{7}\to \Sigma_{7}+\Sigma_1$ & $ 23.356 $ \\ 
 $T_{7}\to \Sigma_{5}+\Sigma_{2}$  & $ 117.12 $ & 
 $T_{7}\to    \Op_{5}+   \Op_{2}$  & $ 0.446 $ & 
 & \\ 
 $T_{7}\to \Sigma_{4}+\Sigma_{3}$  & $ 124.22 $ & 
 $T_{7}\to    \Op_{4}+   \Op_{3}$  & $ 0.552 $ & 
 & \\ 
\hline 
 $T_{8}\to \Sigma_{7}+\Sigma_{1}$  & $ 94.173 $ & 
 $T_{8}\to    \Op_{7}+   \Op_{1}$  & $ 0.223 $ & 
$T_{8}\to \Sigma_{8}+\Sigma_1$ & $ 21.206 $ \\ 
 $T_{8}\to \Sigma_{6}+\Sigma_{2}$  & $ 115.91 $ & 
 $T_{8}\to    \Op_{6}+   \Op_{2}$  & $ 0.467 $ & 
 & \\ 
 $T_{8}\to \Sigma_{5}+\Sigma_{3}$  & $ 124.11 $ & 
 $T_{8}\to    \Op_{5}+   \Op_{3}$  & $ 0.599 $ & 
 & \\ 
 $T_{8}\to \Sigma_{4}+\Sigma_{4}$  & $ 126.37 $ & 
 $T_{8}\to    \Op_{4}+   \Op_{4}$  & $ 0.639 $ & 
 & \\ 
\hline 
 $T_{9}\to \Sigma_{8}+\Sigma_{1}$  & $ 92.918 $ & 
 $T_{9}\to    \Op_{8}+   \Op_{1}$  & $ 0.227 $ & 
$T_{9}\to \Sigma_{9}+\Sigma_1$ & $ 19.6 $ \\ 
 $T_{9}\to \Sigma_{7}+\Sigma_{2}$  & $ 114.78 $ & 
 $T_{9}\to    \Op_{7}+   \Op_{2}$  & $ 0.482 $ & 
 & \\ 
 $T_{9}\to \Sigma_{6}+\Sigma_{3}$  & $ 123.61 $ & 
 $T_{9}\to    \Op_{6}+   \Op_{3}$  & $ 0.631 $ & 
 & \\ 
 $T_{9}\to \Sigma_{5}+\Sigma_{4}$  & $ 127.07 $ & 
 $T_{9}\to    \Op_{5}+   \Op_{4}$  & $ 0.697 $ & 
 & \\ 
\hline 
 $T_{10}\to \Sigma_{9}+\Sigma_{1}$  & $ 91.875 $ & 
 $T_{10}\to    \Op_{9}+   \Op_{1}$  & $ 0.23 $ & 
$T_{10}\to \Sigma_{10}+\Sigma_1$ & $ 18.358 $ \\ 
 $T_{10}\to \Sigma_{8}+\Sigma_{2}$  & $ 113.78 $ & 
 $T_{10}\to    \Op_{8}+   \Op_{2}$  & $ 0.493 $ & 
 & \\ 
 $T_{10}\to \Sigma_{7}+\Sigma_{3}$  & $ 122.98 $ & 
 $T_{10}\to    \Op_{7}+   \Op_{3}$  & $ 0.654 $ & 
 & \\ 
 $T_{10}\to \Sigma_{6}+\Sigma_{4}$  & $ 127.14 $ & 
 $T_{10}\to    \Op_{6}+   \Op_{4}$  & $ 0.738 $ & 
 & \\ 
 $T_{10}\to \Sigma_{5}+\Sigma_{5}$  & $ 128.37 $ & 
 $T_{10}\to    \Op_{5}+   \Op_{5}$  & $ 0.764 $ & 
 & \\ 
\hline 
\end{tabular} 
\\[12pt]
\textbf{Table 1}: The unpolarized decay amplitudes for the glueballs
    $T_k$ with $k\le10$.

\vfill

\begin{tabular}{|c|r||c|r||c|r|} 
\hline 
 & & & &  $\Op_{2}\to \Sigma_{2}+\Sigma_1$ & $ 64.0 $ \\ 
\hline 
 $\Op_{3}\to \Sigma_{2}+\Sigma_{1}$  & $ 3.6571 $ & 
 $\Op_{3}\to \Op_{1}+T_{1}$  & $ 0.163 $ & 
$\Op_{3}\to \Sigma_{3}+\Sigma_1$ & $ 100.0 $ \\ 
\hline 
 $\Op_{4}\to \Sigma_{3}+\Sigma_{1}$  & $ 8.5714 $ & 
 $\Op_{4}\to \Op_{1}+T_{2}$  & $ 0.213 $ & 
$\Op_{4}\to \Sigma_{4}+\Sigma_1$ & $ 116.64 $ \\ 
 $\Op_{4}\to \Sigma_{2}+\Sigma_{2}$  & $ 39.184 $ & 
 $\Op_{4}\to \Op_{2}+T_{1}$  & $ 0.302 $ & 
 & \\ 
\hline 
 $\Op_{5}\to \Sigma_{4}+\Sigma_{1}$  & $ 12.567 $ & 
 $\Op_{5}\to \Op_{1}+T_{3}$  & $ 0.233 $ & 
$\Op_{5}\to \Sigma_{5}+\Sigma_1$ & $ 125.44 $ \\ 
 $\Op_{5}\to \Sigma_{3}+\Sigma_{2}$  & $ 81.455 $ & 
 $\Op_{5}\to \Op_{2}+T_{2}$  & $ 0.422 $ & 
 & \\ 
 & $  $ & 
 $\Op_{5}\to \Op_{3}+T_{1}$  & $ 0.374 $ & 
 & \\ 
\hline 
 $\Op_{6}\to \Sigma_{5}+\Sigma_{1}$  & $ 15.664 $ & 
 $\Op_{6}\to \Op_{1}+T_{4}$  & $ 0.242 $ & 
$\Op_{6}\to \Sigma_{6}+\Sigma_1$ & $ 130.61 $ \\ 
 $\Op_{6}\to \Sigma_{4}+\Sigma_{2}$  & $ 118.67 $ & 
 $\Op_{6}\to \Op_{2}+T_{3}$  & $ 0.479 $ & 
 & \\ 
 $\Op_{6}\to \Sigma_{3}+\Sigma_{3}$  & $ 178.0 $ & 
 $\Op_{6}\to \Op_{3}+T_{2}$  & $ 0.541 $ & 
 & \\ 
 & $  $ & 
 $\Op_{6}\to \Op_{4}+T_{1}$  & $ 0.413 $ & 
 & \\ 
\hline 
 $\Op_{7}\to \Sigma_{6}+\Sigma_{1}$  & $ 18.085 $ & 
 $\Op_{7}\to \Op_{1}+T_{5}$  & $ 0.247 $ & 
$\Op_{7}\to \Sigma_{7}+\Sigma_1$ & $ 133.9 $ \\ 
 $\Op_{7}\to \Sigma_{5}+\Sigma_{2}$  & $ 149.61 $ & 
 $\Op_{7}\to \Op_{2}+T_{4}$  & $ 0.509 $ & 
 & \\ 
 $\Op_{7}\to \Sigma_{4}+\Sigma_{3}$  & $ 271.35 $ & 
 $\Op_{7}\to \Op_{3}+T_{3}$  & $ 0.626 $ & 
 & \\ 
 & $  $ & 
 $\Op_{7}\to \Op_{4}+T_{2}$  & $ 0.61 $ & 
 & \\ 
 & $  $ & 
 $\Op_{7}\to \Op_{5}+T_{1}$  & $ 0.437 $ & 
 & \\ 
\hline 
 $\Op_{8}\to \Sigma_{7}+\Sigma_{1}$  & $ 20.012 $ & 
 $\Op_{8}\to \Op_{1}+T_{6}$  & $ 0.25 $ & 
$\Op_{8}\to \Sigma_{8}+\Sigma_1$ & $ 136.11 $ \\ 
 $\Op_{8}\to \Sigma_{6}+\Sigma_{2}$  & $ 175.15 $ & 
 $\Op_{8}\to \Op_{2}+T_{5}$  & $ 0.526 $ & 
 & \\ 
 $\Op_{8}\to \Sigma_{5}+\Sigma_{3}$  & $ 354.58 $ & 
 $\Op_{8}\to \Op_{3}+T_{4}$  & $ 0.674 $ & 
 & \\ 
 $\Op_{8}\to \Sigma_{4}+\Sigma_{4}$  & $ 429.65 $ & 
 $\Op_{8}\to \Op_{4}+T_{3}$  & $ 0.716 $ & 
 & \\ 
 & $  $ & 
 $\Op_{8}\to \Op_{5}+T_{2}$  & $ 0.654 $ & 
 & \\ 
 & $  $ & 
 $\Op_{8}\to \Op_{6}+T_{1}$  & $ 0.452 $ & 
 & \\ 
\hline 
 $\Op_{9}\to \Sigma_{8}+\Sigma_{1}$  & $ 21.576 $ & 
 $\Op_{9}\to \Op_{1}+T_{7}$  & $ 0.252 $ & 
$\Op_{9}\to \Sigma_{9}+\Sigma_1$ & $ 137.67 $ \\ 
 $\Op_{9}\to \Sigma_{7}+\Sigma_{2}$  & $ 196.36 $ & 
 $\Op_{9}\to \Op_{2}+T_{6}$  & $ 0.537 $ & 
 & \\ 
 $\Op_{9}\to \Sigma_{6}+\Sigma_{3}$  & $ 427.02 $ & 
 $\Op_{9}\to \Op_{3}+T_{5}$  & $ 0.704 $ & 
 & \\ 
 & $  $ & 
 $\Op_{9}\to \Op_{4}+T_{4}$  & $ 0.779 $ & 
 & \\ 
 & $  $ & 
 $\Op_{9}\to \Op_{5}+T_{3}$  & $ 0.774 $ & 
 & \\ 
 & $  $ & 
 $\Op_{9}\to \Op_{6}+T_{2}$  & $ 0.683 $ & 
 & \\ 
 & $  $ & 
 $\Op_{9}\to \Op_{7}+T_{1}$  & $ 0.463 $ & 
 & \\ 
\hline 
 $\Op_{10}\to \Sigma_{9}+\Sigma_{1}$  & $ 22.867 $ & 
 $\Op_{10}\to \Op_{1}+T_{8}$  & $ 0.253 $ & 
$\Op_{10}\to \Sigma_{10}+\Sigma_1$ & $ 138.81 $ \\ 
 $\Op_{10}\to \Sigma_{8}+\Sigma_{2}$  & $ 214.15 $ & 
 $\Op_{10}\to \Op_{2}+T_{7}$  & $ 0.544 $ & 
 & \\ 
 & $  $ & 
 $\Op_{10}\to \Op_{3}+T_{6}$  & $ 0.723 $ & 
 & \\ 
 & $  $ & 
 $\Op_{10}\to \Op_{4}+T_{5}$  & $ 0.818 $ & 
 & \\ 
 & $  $ & 
 $\Op_{10}\to \Op_{5}+T_{4}$  & $ 0.847 $ & 
 & \\ 
 & $  $ & 
 $\Op_{10}\to \Op_{6}+T_{3}$  & $ 0.814 $ & 
 & \\ 
 & $  $ & 
 $\Op_{10}\to \Op_{7}+T_{2}$  & $ 0.703 $ & 
 & \\ 
 & $  $ & 
 $\Op_{10}\to \Op_{8}+T_{1}$  & $ 0.47 $ & 
 & \\ 
\hline 
\end{tabular} 
\\[12pt]
\textbf{Table 2}: The unpolarized decay amplitudes for the glueballs
    $\Op_k$ with $k\le10$.

\vfill

\begin{tabular}{|c|l||c|l||c|l|} 
\hline 
 & &  
$\Sigma_{2}\to    T_{1}+\Sigma_1$ & $ 0.0 $ &  
$\Sigma_{2}\to \Op_{1}+\Sigma_1$ & $ 115.2 $ \\ 
\hline 
 $\Sigma_{3}\to T_{1}+\Sigma_{1}$  & $ 29.623 $ & 
$\Sigma_{3}\to    T_{2}+\Sigma_1$ & $ 0.091 $ &  
$\Sigma_{3}\to \Op_{2}+\Sigma_1$ & $ 91.428 $ \\ 
\hline 
 $\Sigma_{4}\to T_{1}+\Sigma_{2}$  & $ 35.265 $ & 
$\Sigma_{4}\to    T_{3}+\Sigma_1$ & $ 0.49 $ &  
$\Sigma_{4}\to \Op_{3}+\Sigma_1$ & $ 77.143 $ \\ 
 $\Sigma_{4}\to T_{2}+\Sigma_{1}$  & $ 44.082 $ & 
 & & & \\ 
\hline 
 $\Sigma_{5}\to T_{1}+\Sigma_{3}$  & $ 37.403 $ & 
$\Sigma_{5}\to    T_{4}+\Sigma_1$ & $ 1.0473 $ &  
$\Sigma_{5}\to \Op_{4}+\Sigma_1$ & $ 68.422 $ \\ 
 $\Sigma_{5}\to T_{2}+\Sigma_{2}$  & $ 53.432 $ & 
 & & & \\ 
 $\Sigma_{5}\to T_{3}+\Sigma_{1}$  & $ 52.364 $ & 
 & & & \\ 
\hline 
 $\Sigma_{6}\to T_{1}+\Sigma_{4}$  & $ 38.449 $ & 
$\Sigma_{6}\to    T_{5}+\Sigma_1$ & $ 1.6275 $ &  
$\Sigma_{6}\to \Op_{5}+\Sigma_1$ & $ 62.657 $ \\ 
 $\Sigma_{6}\to T_{2}+\Sigma_{3}$  & $ 57.215 $ & 
 & & & \\ 
 $\Sigma_{6}\to T_{3}+\Sigma_{2}$  & $ 64.081 $ & 
 & & & \\ 
 $\Sigma_{6}\to T_{4}+\Sigma_{1}$  & $ 57.673 $ & 
 & & & \\ 
\hline 
 $\Sigma_{7}\to T_{1}+\Sigma_{5}$  & $ 39.04 $ & 
$\Sigma_{7}\to    T_{6}+\Sigma_1$ & $ 2.1737 $ &  
$\Sigma_{7}\to \Op_{6}+\Sigma_1$ & $ 58.595 $ \\ 
 $\Sigma_{7}\to T_{2}+\Sigma_{4}$  & $ 59.152 $ & 
 & & & \\ 
 $\Sigma_{7}\to T_{3}+\Sigma_{3}$  & $ 69.011 $ & 
 & & & \\ 
 $\Sigma_{7}\to T_{4}+\Sigma_{2}$  & $ 70.982 $ & 
 & & & \\ 
 $\Sigma_{7}\to T_{5}+\Sigma_{1}$  & $ 61.349 $ & 
 & & & \\ 
\hline 
 $\Sigma_{8}\to T_{1}+\Sigma_{6}$  & $ 39.408 $ & 
$\Sigma_{8}\to    T_{7}+\Sigma_1$ & $ 2.6682 $ &  
$\Sigma_{8}\to \Op_{7}+\Sigma_1$ & $ 55.588 $ \\ 
 $\Sigma_{8}\to T_{2}+\Sigma_{5}$  & $ 60.283 $ & 
 & & & \\ 
 $\Sigma_{8}\to T_{3}+\Sigma_{4}$  & $ 71.609 $ & 
 & & & \\ 
 $\Sigma_{8}\to T_{4}+\Sigma_{3}$  & $ 76.724 $ & 
 & & & \\ 
 $\Sigma_{8}\to T_{5}+\Sigma_{2}$  & $ 75.784 $ & 
 & & & \\ 
 $\Sigma_{8}\to T_{6}+\Sigma_{1}$  & $ 64.038 $ & 
 & & & \\ 
\hline 
 $\Sigma_{9}\to T_{1}+\Sigma_{7}$  & $ 39.652 $ & 
$\Sigma_{9}\to    T_{8}+\Sigma_1$ & $ 3.1094 $ &  
$\Sigma_{9}\to \Op_{8}+\Sigma_1$ & $ 53.278 $ \\ 
 $\Sigma_{9}\to T_{2}+\Sigma_{6}$  & $ 61.003 $ & 
 & & & \\ 
 $\Sigma_{9}\to T_{3}+\Sigma_{5}$  & $ 73.161 $ & 
 & & & \\ 
 $\Sigma_{9}\to T_{4}+\Sigma_{4}$  & $ 79.812 $ & 
 & & & \\ 
 $\Sigma_{9}\to T_{5}+\Sigma_{3}$  & $ 82.119 $ & 
 & & & \\ 
 $\Sigma_{9}\to T_{6}+\Sigma_{2}$  & $ 79.303 $ & 
 & & & \\ 
 $\Sigma_{9}\to T_{7}+\Sigma_{1}$  & $ 66.086 $ & 
 & & & \\ 
\hline 
 $\Sigma_{10}\to T_{1}+\Sigma_{8}$  & $ 39.822 $ & 
$\Sigma_{10}\to    T_{9}+\Sigma_1$ & $ 3.5011 $ &  
$\Sigma_{10}\to \Op_{9}+\Sigma_1$ & $ 51.45 $ \\ 
 $\Sigma_{10}\to T_{2}+\Sigma_{7}$  & $ 61.49 $ & 
 & & & \\ 
 $\Sigma_{10}\to T_{3}+\Sigma_{6}$  & $ 74.167 $ & 
 & & & \\ 
 $\Sigma_{10}\to T_{4}+\Sigma_{5}$  & $ 81.687 $ & 
 & & & \\ 
 $\Sigma_{10}\to T_{5}+\Sigma_{4}$  & $ 85.577 $ & 
 & & & \\ 
 $\Sigma_{10}\to T_{6}+\Sigma_{3}$  & $ 86.086 $ & 
 & & & \\ 
 $\Sigma_{10}\to T_{7}+\Sigma_{2}$  & $ 81.987 $ & 
 & & & \\ 
 $\Sigma_{10}\to T_{8}+\Sigma_{1}$  & $ 67.698 $ & 
 & & & \\ 
\hline 
\end{tabular} 
\\[12pt]
\textbf{Table 3}: The unpolarized decay amplitudes for the glueballs
    $\Sigma_k$ with $k\le10$.
\end{center}

\vfill

\end{appendix}

\bibliographystyle{JHEP}
\bibliography{GPPZ_refs}

\end{document}